\documentclass[%
 reprint,
 amsmath,amssymb,
 aps,
prstab,
]{revtex4-2}

\usepackage{graphicx}
\usepackage{dcolumn}
\usepackage{bm}

\usepackage{siunitx}
\usepackage{xcolor}
\usepackage{braket}
\usepackage{mathtools}
\usepackage{amsmath}
\usepackage{leftidx}
\usepackage{wasysym}
\usepackage{float}

\newcommand{\N}{\mathcal{N}}

\newcommand{\NPhc}{\N_{\mathrm{c}}}

\newcommand{\NPhone}{\N_{\mathrm{c}}^{(1)}}
\newcommand{\abs}[1]{\left| #1 \right|}
\newcommand{\av}[1]{\langle #1 \rangle}

\newcommand{\Rm}{\bm{r}_m}

\newcommand{\var}[1]{\mathrm{var}(#1)}
\newcommand{\nbar}{\av{\NPhc}}
\newcommand{\kv}{\bm{k}}
\newcommand{\kvone}{\kv_1}
\newcommand{\kvtwo}{\kv_2}
\newcommand{\dkv}{d\bm{k}}
\newcommand{\dkvone}{d\kv_1}
\newcommand{\dkvtwo}{d\kv_2}
\newcommand{\If}[1]{I^{(1)}_{#1}}
\newcommand{\I}{\If{\kv}}
\newcommand{\Itot}{I_{\kv}}
\newcommand{\Ione}{\If{\kvone}}
\newcommand{\Itwo}{\If{\kvtwo}}
\newcommand{\none}{\bm{n}_1}
\newcommand{\ntwo}{\bm{n}_2}
\newcommand{\tonex}{\theta_{1x}}
\newcommand{\ttwox}{\theta_{2x}}
\newcommand{\toney}{\theta_{1y}}
\newcommand{\ttwoy}{\theta_{2y}}
\newcommand{\sx}{\sigma_x}
\newcommand{\sy}{\sigma_y}
\newcommand{\sz}{\sigma_z}

\newcommand{\K}{\bm{\mathrm{K}}}
\newcommand{\Kone}{\K_1}
\newcommand{\Ktwo}{\K_2}
\newcommand{\Ss}{\bm{\Sigma}}
\newcommand{\Konetwo}{\K_{12}}
\newcommand{\kx}{k_x}
\newcommand{\ky}{k_y}
\newcommand{\kz}{k_z}
\newcommand{\Donetwo}{\bm{\Delta}_{12}}
\newcommand{\alphak}{\alpha_{\kv}}
\newcommand{\nk}{n_{\kv}}
\newcommand{\nkop}{\hat{n}_{\kv}}
\newcommand{\qef}[1]{\eta_{#1}}
\newcommand{\qe}{\qef{\kv}}
\newcommand{\qeone}{\qef{\kvone}}
\newcommand{\qetwo}{\qef{\kvtwo}}
\newcommand{\ak}{\hat{a}_{\kv}}
\newcommand{\bk}{\hat{b}_{\kv}}
\newcommand{\vk}{\hat{v}_{\kv}}
\newcommand{\dk}{\hat{d}_{\kv}}
\newcommand{\Nk}{\N_{\kv}}
\newcommand{\Ne}{N_e}
\newcommand{\avNqc}{\av{\NPhc}}
\newcommand{\ct}{\kappa}

\newcommand{\Cf}{C_{\mathrm{f}}}
\newcommand{\Rf}{R_{\mathrm{f}}}
\newcommand{\Nu}{N_\mathrm{u}}
\newcommand{\D}{$\Delta$}
\renewcommand{\S}{$\Sigma$}
\newcommand{\Ib}{I_{\mathrm{beam}}}
\newcommand{\ex}{\epsilon_x}
\newcommand{\ey}{\epsilon_y}
\newcommand{\sE}{\sigma_p}

\begin{document}

\title{Statistical properties of undulator radiation in the IOTA storage ring}

\author{Ihar Lobach}
\email{ilobach@uchicago.edu}
\affiliation{The University of Chicago, Chicago, IL 60637, USA}

\author{Valeri Lebedev}
\author{Sergei Nagaitsev}
\altaffiliation[Also at ]{the University of Chicago.}
\author{Aleksandr Romanov}
\author{Giulio Stancari} 
\author{Alexander Valishev} 
\affiliation{Fermi National Accelerator Laboratory, Batavia, IL 60510, USA}%


\author{Aliaksei Halavanau}
\author{Zhirong Huang}
\affiliation{SLAC National Accelerator Laboratory, Stanford University, Menlo Park CA 94025, USA}%

\author{Kwang-Je Kim}
\altaffiliation[Also at ]{the University of Chicago.}
\affiliation{Argonne National Accelerator Laboratory, Lemont, IL 60439, USA}%



\begin{abstract}
We study turn-by-turn fluctuations in the number of spontaneously emitted photons from an undulator, installed in the Integrable Optics Test Accelerator (IOTA) electron storage ring at Fermilab. A theoretical model is presented, showing the relative contributions due to the discrete nature of light emission and to the incoherent sum of fields from different electrons in the bunch. The model is compared with a previous experiment at Brookhaven and with new experiments we carried out at IOTA. Our experiments focused on the case of a large number of longitudinal and transverse radiation modes, a regime where photon shot noise is significant and the total magnitude of the fluctuations is very small. The experimental and data analysis techniques, required to reach the desired sensitivity, are detailed. We discuss how the model and the experiment provide insights into this emission regime, enable diagnostics of small beam sizes, and improve our understanding of beam lifetime in IOTA.

\end{abstract}

\maketitle


\section{\label{sec:level1}Introduction}
In the last few decades there were several experiments regarding statistical properties of incoherent synchrotron radiation, produced by electron bunches in storage rings and linear accelerators \cite{teich1990statistical,sannibale2009absolute,catravas1999measurement,sajaev2004measurement,sajaev2000determination}. The fluctuation in the radiated energy (or the number of photons) from pulse to pulse was studied experimentally and theoretically. It was shown in \cite{sannibale2009absolute,catravas1999measurement} that in some cases the rms electron bunch length can be measured via this fluctuation. Moreover, references \cite{sajaev2004measurement,sajaev2000determination} suggest that if the fluctuations in radiation spectrum are measured with a high resolution spectrometer, then even the shape of the electron bunch can be reconstructed. These observations, combined with the fact that fluctuations of the same nature are present in SASE FELs \cite{kim1997start,kim2017synchrotron,benson1985shot,saldin2013physics,pellegrini2016physics,becker1982photon}, make the study of fluctuations in incoherent synchrotron radiation relevant for the understanding of beam dynamics and for beam diagnostics.

The number of photons radiated incoherently by an electron bunch in an external electromagnetic field (undulator, wiggler, dipole magnet, etc.) fluctuates from pass to pass due to the following two mechanisms \cite{goodman2015statistical}. The first mechanism is the photon shot noise, related to the quantum discrete nature of light. This effect would exist even if there was only one electron. Indeed, the electron would radiate light with Poisson statistics \cite{glauber1963quantum,glauber1963coherent,glauber1951some}. The second mechanism is due to the fact that the field produced by a bunch of electrons is an incoherent sum of fields produced by all the electrons in the bunch. If wave packets of radiation produced by different electrons overlap, the incoherent sum fluctuates from pass to pass because the positions of the electrons in the bunch change. In a storage ring, the effect arises because of betatron motion, synchrotron motion, radiation induced diffusion, etc.; in linacs, assuming exactly equal bunch charges, there is a new bunch of electrons at every pass, the positions of which are not correlated with the positions of electrons in the previous bunch.

For dense bunches, the fluctuations in the number of emitted photons are usually dominated by the incoherence contribution \cite{kim2017synchrotron}, as it was the case in  \cite{teich1990statistical,sajaev2004measurement,sajaev2000determination,sannibale2009absolute,catravas1999measurement}. In this paper, we present the results of studies of statistical properties of undulator radiation in the IOTA ring at Fermilab \cite{antipov2017iota}, where the contributions from both mechanisms are comparable. This also means that the fluctuations were very small (compared to \cite{teich1990statistical}, for instance), and we present several critical improvements to the setup from \cite{teich1990statistical,sannibale2009absolute}.

We start by reviewing the theory of fluctuations in synchrotron radiation. The theory is relevant for both storage rings and linear accelerators. However, below we assume a radiation coming from a single bunch in a storage ring, where the number of participating electrons does not fluctuate. 
We derive an equation for the variance of the number of detected photons for a Gaussian electron bunch, taking into account the quantum efficiency of the detector. Then, the theoretical predictions are compared with the empirical data from \cite{teich1990statistical}. Finally, we describe our experiment in IOTA \cite{lobach2019study}, and how the empirical data from this experiment compare with our model.

\section{Theoretical model}
Consider incoherent synchrotron radiation (undulator, wiggler, bending-magnet, etc.), emitted by a Gaussian bunch consisting of many randomly located electrons. Let us assume that the synchrotron radiation is collected in a wide spectral range, and also in a large solid angle.

\subsection{Quantum fluctuations}\label{subsec:quantum_fluctuations}
Whereas in general we consider a bunch with randomly located electrons, in this subsection we fix positions (phases) of all the electrons and derive the quantum contribution to the fluctuations. We take an ensemble average over random positions (phases) of the electrons in subsection~\ref{subsec:classical_fluctuations}.

Conceptually, one can divide the detector into many detectors, each sensing only one mode of the produced radiation with the wave vector $\kv$. The volume of the $\kv$-space associated with this mode will be denoted by $\dkv \equiv dk_xdk_ydk_z$. We can consider periodic boundary conditions in a cube with side $L$, then $d\kv = (2\pi/L)^3$. We will always take the limit $L\rightarrow\infty$ at some point, and all the sums over modes $\kv$ will be replaced by integrals. Upon this transition to integrals, it will also be valid to use $d\kv=k^2dkd\Omega$, where $\Omega$ is the solid angle.

It was shown in \cite{glauber1963coherent,glauber1963quantum} that any classical current (corresponding to a negligible electron recoil) produces a coherent state of radiated electromagnetic field. The coherent state in the single mode $\kv$ is given by \cite{glauber1963coherent}
\begin{equation}\label{eq:coh_state_alpha_k}
    \ket{\alphak} = e^{-\frac{1}{2}\abs{\alphak}^2}\sum_{\nk}\frac{\alphak^{\nk}}{\sqrt{\nk!}}\ket{\nk},
\end{equation}
\noindent where $\ket{\nk}$ is the state with $\nk$ photons in the mode $\kv$ and the exact formula for $\alphak$ is provided in \cite{glauber1963coherent}. For our purposes, it is sufficient to mention that
\begin{equation}\label{eq:Idef}
    \abs{\alphak}^2=\NPhc^{\kv,d\kv}=\Itot\dkv,
\end{equation}
\noindent where $\NPhc^{\kv,d\kv}$ is the quasi-classical number of photons emitted in the mode $\kv$, and $\Itot$ is the quasi-classical spectral-angular density of the number of emitted photons, defined in this paper by the second equality in Eq.~\eqref{eq:Idef}. There are two orthogonal polarization components for each wavevector $\kv$, usually denoted by $\sigma$ and $\pi$ \cite{kim2017synchrotron}. In this paper, we omit $\sigma$ and $\pi$ for the sake of brevity. If there is a sum or an integral over $\kv$ in any equation, it should be understood that there is a sum over both polarizations as well.

In the coherent state Eq.~\eqref{eq:coh_state_alpha_k}, the number of photons $\nk$ obeys the Poisson statistics. In fact, the mean and the variance of $\nk$ are equal and given by $\abs{\alphak}^2=\Itot d\kv$:
\begin{align}
\av{\nk}=\braket{\alphak|\nkop|\alphak}=\abs{\alphak}^2=\Itot d\kv,\label{eq:avnk}\\
\var{\nk}=\braket{\alphak|(\nkop-\av{\nk})^2|\alphak}=\abs{\alphak}^2=\Itot\dkv\label{eq:varnk},
\end{align}
\noindent where $\nkop=\ak^\dagger\ak$, with $\ak^\dagger$ and $\ak$ being the creation and annihilation operators for the mode $\kv$.

\begin{figure}[!h]
   \centering
   \includegraphics*[width=0.75\columnwidth]{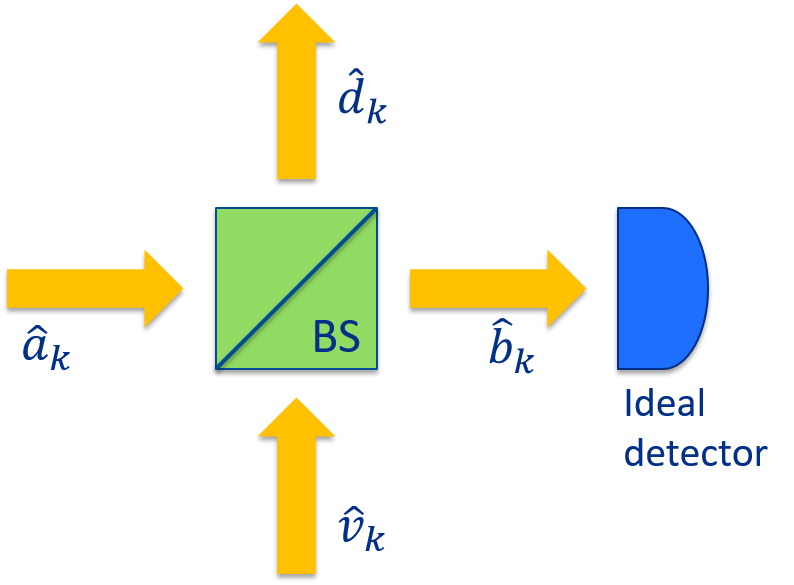}
   \caption{The beam splitter model for quantum efficiency of a non-ideal detector.}
   \label{fig:BS_QE_model}
\end{figure}

A conventional approach will be used to account for the $\kv$-dependent quantum efficiency of the detector, $\qe$, namely, the beam splitter model, described, for example, in \cite{meda2017photon}, also see Fig.~\ref{fig:BS_QE_model}.

The input-output relations for the beam splitter take the form
\begin{align}
\bk=\sqrt{\qe}\ak+i\sqrt{1-\qe}\vk,\label{eq:bk}\\
\dk=\sqrt{1-\qe}\vk+i\sqrt{\qe}\ak,
\end{align}
\noindent where $\ak$ is the incoming field, $\vk$ corresponds to the second input port, which is in the vacuum state in this model, $\bk$ and $\dk$ are transmitted and reflected fields, respectively. Equation~\eqref{eq:bk} lets us calculate the mean and the variance for the number of detected photons $\Nk$ \cite{meda2017photon}
\begin{equation}\label{eq:avNk}
    \av{\Nk}=\leftidx{_{a,v}}{\braket{\alphak,0|\bk^\dagger\bk|\alphak,0}}{_{a,v}} = \qe\av{\nk},
\end{equation}
\begin{multline}\label{eq:varNk}
    \var{\Nk}=\leftidx{_{a,v}}{\braket{\alphak,0|(\bk^\dagger\bk-\av{\Nk})^2|\alphak,0}}{_{a,v}}=\\
    \qe\av{\nk}+\qe^2(\var{\nk}-\av{\nk}).
\end{multline}

In the second term in Eq.~\eqref{eq:varNk} one can see that it is important to have a high quantum efficiency to be able to observe the sub- or super-Poisson statistics \cite{chen1999photon,chen2001observation,madey2014wilson} of a quantum origin. However, for coherent states this term vanishes, and, using Eqs.~\eqref{eq:avnk} and \eqref{eq:varnk}, we obtain
\begin{equation}
    \av{\Nk}=\var{\Nk}=\qe \Itot d\kv.
\end{equation}

In fact, using Eq.~\eqref{eq:bk} to find how the coherent state Eq.~\eqref{eq:coh_state_alpha_k} is transformed by the beam splitter, one can show that the output state will be a coherent state, and the number of detected photons $\Nk$ will obey the Poisson statistics as well as $\nk$ does. Furthermore,  since a sum of independent random Poisson variables is Poissonian \cite{grimmett2014probability}, the total number of detected photons $\N$,
\begin{equation}
    \N=\sum\limits_{\kv}\Nk,
\end{equation}
\noindent will obey the Poisson distribution with the mean and the variance given by
\begin{equation}\label{eq:avN}
    \av{\N}=\var{\N}=\int\qe\Itot d\kv=\NPhc,
\end{equation}
\noindent where $\NPhc$ is the total number of detected photons, calculated in the quasi-classical model, the integration is performed over all $\kv$. In Eq.~\eqref{eq:avN} and below in this paper,  $\av{\N}$ denotes the average of the random variable $\N$, not to be confused with $\av{\nk}$ in Eq.~\eqref{eq:avnk} or $\av{\Nk}$ in Eq.~\eqref{eq:avNk}, where it denotes the expectation values of quantum operators $\ak^\dagger\ak$ and $\bk^\dagger\bk$ in corresponding coherent states. All sources of losses can be incorporated into $\qe$, such as the detector's quantum efficiency, losses in focusing lenses (if used), in the vacuum chamber windows, or due to the finite detector acceptance ($\qe$ can be set to zero for radiation angles that are not collected by the detector). If a spectral filter is used, its transmission function can be incorporated into $\qe$ and polarization filters can be accounted for in a similar manner.

In \cite{glauber1951some}, it was shown that the total number of photons emitted (not detected) by a classical current obeys the Poisson distribution. Equation~\eqref{eq:avN} is in agreement with this result, and extends it to the counts in a non-ideal detector. It is also noteworthy that for $\av{\N}\gg 1$ the distribution for $\N$ is essentially Gaussian.

Equation~\eqref{eq:avN} is in agreement with intuitive understanding and is usually taken for granted. In the derivations presented above, we merely reviewed the proof of Eq.~\eqref{eq:avN} for synchrotron radiation from the quantum optics perspective.

\subsection{Classical fluctuations}\label{subsec:classical_fluctuations}



The results of the previous subsection can be summarized as follows
\begin{equation}\label{eq:instance_of_n}
    \N=\Delta\left(\NPhc\right)+\NPhc,
\end{equation}
\noindent where $\N$ is the number of photons detected at a certain turn; $\NPhc$ is the quasi-classical prediction for the number of detected photons produced by an electron bunch with fixed phases; $\Delta(\NPhc)$ is a random variable describing the quantum fluctuation around the mean value $\NPhc$. From Eq.~\eqref{eq:avN} it follows that
\begin{align}
    \av{\Delta\left(\NPhc\right)}=0,\label{eq:avDeltaNPhs}\\
    \var{\Delta\left(\NPhc\right)}=\NPhc.\label{eq:varDeltaNPhc}
\end{align}

In this subsection we take into account the fact that in reality in a storage ring electron phases are random each turn. Hence, $\NPhc$ changes as well. One can still use Eq.~\eqref{eq:instance_of_n} to represent the number of detected photons per turn, and we will use the same symbol for this variable $\N$ in this subsection. However, it should be understood that the random variable $\N$ in this subsection is different from that in Subsec.~\ref{subsec:quantum_fluctuations}. In this subsection, $\N$ incorporates both the quantum fluctuation $\Delta\left(\NPhc\right)$ and the classical fluctuation of $\NPhc$ from turn to turn.

One can calculate $\var{\N}$ by taking the variance of Eq.~\eqref{eq:instance_of_n}:
\begin{equation}\label{eq:varNsum}
    \var{\N}=\var{\Delta\left(\NPhc\right)+\NPhc}.
\end{equation}

We would like to use the fact that 
\begin{equation}
    \var{a+b}=\var{a}+\var{b}
\end{equation}
\noindent  for independent $a$ and $b$. However, $\Delta\left(\NPhc\right)$ and $\NPhc$ are, in general, not independent. Nonetheless, if, on average, a large number of photons is detected per turn $\av{\NPhc}\gg1$, then $\Delta\left(\NPhc\right)$ can be approximated by $\Delta\left(\av{\NPhc}\right)$, which is independent of $\NPhc$. In this approximation, Eq.~\eqref{eq:varNsum} becomes
\begin{equation}\label{eq:varNasSum}
    \var{\N}=\var{\Delta\left(\av{\NPhc}\right)}+\var{\NPhc}.
\end{equation}

Equation~\eqref{eq:varDeltaNPhc} can be used in Eq.~\eqref{eq:varNasSum} to obtain
\begin{equation}\label{eq:varN1}
    \var{\N}=\av{\NPhc}+\var{\NPhc},
\end{equation}
\noindent where $\av{\NPhc}$ can be replaced with $\av{\N}$ according to Eqs.~\eqref{eq:instance_of_n} and \eqref{eq:avDeltaNPhs}.

Equation~\eqref{eq:varN1} can be rewritten in the form of \cite{kim2017synchrotron,kim1997temporal}
\begin{equation}\label{eq:varN_from_book}
    \var{\N}=\av{\N}+\frac{1}{M}\av{\N}^2,
\end{equation}
\noindent where the parameter $M$ was introduced. In this paper it is defined as
\begin{equation}\label{eq:defM}
    \frac{1}{M}\equiv\frac{\var{\NPhc}}{\av{\NPhc}^2},
\end{equation}
\noindent however, it can be identified with the number of coherent modes defined in \cite{kim2017synchrotron,kim1989characteristics}, therefore we will use this name for the parameter $M$ from now on.

In this subsection, we find $M$ by explicitly calculating $\var{\NPhc}=\av{\NPhc^2}-\avNqc^2$ and using Eq.~\eqref{eq:defM}. To begin with, we introduce $\I$, i.e., the quasi-classical spectral-angular density of the number of emitted photons for the case when there is only one electron in the storage ring, by the following relation
\begin{equation}
    \frac{d\NPhone}{\dkv}=\qe\I.
\end{equation}

Then, the total quasi-classical number of photons detected in this case is given by
\begin{equation}\label{eq:totalNgamma}
    \NPhone = \int \dkv \qe \I.
\end{equation}

In the derivations below, we will assume that the beam divergence is negligible compared to the radiation divergence \cite{kim2017synchrotron,kim1989characteristics,sannibale2009absolute}
\begin{equation}\label{eq:beam_divergence}
    \sigma_{x'},\sigma_{y'}\ll\sigma_{r'}.
\end{equation}

Also, it is assumed that the electrons' momentum spread $\sE$ is sufficiently small, so that all the electrons in the bunch produce radiation with approximately the same spectrum 
\begin{equation}\label{eq:energy_spread}
    \abs{\frac{\partial\I}{\partial p}}\sE\ll\abs{\I}.
\end{equation}

If conditions \eqref{eq:beam_divergence} and \eqref{eq:energy_spread} are fulfilled (typical for an electron storage ring, \cite{teich1990statistical,lobach2019study,sannibale2009absolute}), one can use the following formula for $\NPhc$ \cite{kim2017synchrotron,kim1989characteristics,sannibale2009absolute}
\begin{equation}\label{eq:nqc}
     \NPhc = \int \dkv\qe \I \abs{\sum\limits_m e^{i\kv\cdot\bm{r}_m}}^2,
\end{equation}
\noindent where $m = 1, \ldots, \Ne$, with $\Ne$ being the number of electrons in the bunch; $\Rm\equiv(x_m,y_m,-ct_m)$,  $x_m$ and $y_m$ describe the transverse position of $m$th electron when it enters the synchrotron light source (undulator, wiggler, bending magnet, etc.) at time $t_m$. Accordingly, the square of $\NPhc$ is given by
\begin{multline}\label{eq:nqc2}
    (\NPhc)^2 = \int \dkvone \dkvtwo\qeone \Ione\qetwo \Itwo \times \\  \abs{\sum\limits_m e^{i\kvone\cdot\bm{r}_m}}^2 \abs{\sum\limits_n e^{i\kvtwo\cdot\bm{r}_n}}^2.
\end{multline}

We consider a Gaussian distribution of particles in the bunch along $x$, $y$ and $ct$, with rms sizes $\sx$, $\sy$ and $\sz$, respectively. The following two mathematical identities can be derived in this case 
\begin{equation}\label{eq:average1}
    \av{\abs{\sum\limits_m e^{i\kv\cdot\Rm}}^2}=\Ne+\Ne\left(\Ne-1\right)e^{-\K\cdot\Ss},
\end{equation}
\begin{widetext}
\begin{multline}\label{eq:average2}
   \av{\abs{\sum\limits_m e^{i\kvone\cdot\bm{r}_m}}^2\abs{\sum\limits_n e^{i\kvtwo\cdot\bm{r}_n}}^2}=
   \Ne^2+\Ne(\Ne-1)e^{-\bm{\Delta}_{12}\cdot\Ss}+\\ \Ne(\Ne-1)\left[\Ne\left(e^{-\Kone\cdot\Ss}+e^{-\Ktwo\cdot\Ss}\right) + 2\left(\Ne-2\right)e^{-\left(\Konetwo+\Donetwo\right)\cdot\Ss}+\left(\Ne^2-3\Ne+3\right)e^{-\left(\Kone+\Ktwo\right)\cdot\Ss}\right],
\end{multline}
\end{widetext}
\noindent where the average is taken over each electron's position, $\K\equiv(\kx^2,\ky^2,\kz^2)$ ($\Kone$ and $\Ktwo$ are defined analogously), $\Ss\equiv(\sx^2,\sy^2,\sz^2)$, $\Konetwo\equiv(k_{1x}k_{2x},k_{1y}k_{2y},k_{1z}k_{2z})$, $\Donetwo\equiv((k_{1x}-k_{2x})^2,(k_{1y}-k_{2y})^2,(k_{1z}-k_{2z})^2)$.

In this paper, we consider radiation with sufficiently short  wavelengths where contribution of coherent synchrotron radiation (CSR) is negligible. It corresponds to the condition where $\kz\sz-\ln{\Ne}\gg 1$. It is satisfied in the Brookhaven experiment \cite{teich1990statistical} ($\kz\sz\sim \SI{7e5}{}$), and in our experiment \cite{lobach2019study} ($\kz\sz\sim \SI{1.2e6}{}$). Then, it is sufficient to keep only the first term in Eq.~\eqref{eq:average1} and the first two terms in Eq.~\eqref{eq:average2}. Hence, it follows from Eqs.~\eqref{eq:nqc} and \eqref{eq:average1} that
\begin{equation}
    \nbar=\Ne\int \dkv \qe \I=\Ne\NPhone.
\end{equation}

Further, assuming that 
\begin{equation}
    \abs{\frac{\partial \I}{\partial \kz}}\frac{1}{\sz}\ll\abs{\I},
\end{equation}
\noindent which is usually fulfilled when $\kz\sz\gg 1$ (bunch length much longer than the radiation wavelength), we can use the following approximation when integrating in Eq.~\eqref{eq:nqc2}
\begin{equation}\label{eq:delta_f_approx}
    e^{-\sz^2(k_{1z}-k_{2z})^2}\sim \frac{\sqrt{\pi}}{\sz}\delta\left(k_{1z}-k_{2z}\right),
\end{equation}
\noindent where $\delta\left(..\right)$ is the Dirac delta function.
  
Keeping only the first two terms in Eq.~\eqref{eq:average2} and using Eq.~\eqref{eq:delta_f_approx} during integration in Eq.~\eqref{eq:nqc2}, and also assuming $\Ne\gg 1$,  we arrive at the following expression for the inverse of the number of coherent modes
\begin{widetext}
\begin{equation}\label{eq:M}
   \frac{1}{M}\equiv \frac{\var{\NPhc}}{\avNqc^2}=\frac{\frac{\sqrt{\pi}}{\sz}\int  dk d\Omega_1 d\Omega_2 k^4  \qef{k\none}\If{k\none}\qef{k\ntwo} \If{k\ntwo} e^{-k^2\sx^2(\tonex-\ttwox)^2-k^2\sy^2(\toney-\ttwoy)^2}}{\left(\int d\kv \qe\I\right)^2},
\end{equation}
\end{widetext}
\noindent where $\none\approx (\tonex,\toney,1)$, $\ntwo\approx (\ttwox,\ttwoy,1)$, i.e., it is assumed that the radiation is concentrated at small angles $\theta_x, \theta_y\lesssim1/\gamma\ll 1$ and the paraxial approximation is used. Equation~\eqref{eq:M} is in agreement with Eq.~(14) of Ref.~\cite{sannibale2009absolute} and with a simple order of magnitude estimate of $M$, see Appendix~\ref{app:Mestimation}. In \cite{sannibale2009absolute}, the authors focus on the model where the electron bunch and spectral-angular distribution of the radiation is assumed to be Gaussian. In our paper, we still consider a Gaussian electron bunch, however, we do not assume the Gaussian spectral-angular distribution for the radiation in Eq.~\eqref{eq:M}. Instead, we have used the expression for spectral-angular intensity distribution for the undulator radiation from Ref. \cite{clarke2004science}.  In the limit of a large transverse electron bunch size, one can use approximations for the $x$- and $y$-directions, analogous to Eq.~\eqref{eq:delta_f_approx}, in Eq.~\eqref{eq:M} to simplify it further. In the opposite limiting case, i.e., $\sx,\sy\rightarrow 0$, one can omit the exponent in Eq.~\eqref{eq:M}. More information on the limiting cases can be found in \cite{kim1995analysis,sannibale2009absolute}. In the numerical examples in this paper (the Brookhaven experiment \cite{teich1990statistical} and our experiment \cite{lobach2019study}) we use the full version of Eq.~\eqref{eq:M} and perform numerical integration, since the values of the parameters in these experiments correspond to an intermediate case. 

The assumption of a Gaussian bunch distribution works well in the IOTA ring for $x$,$x^{\prime}$,$y$,$y^{\prime}$, and $p$. However, in some cases, it does not properly describe the distribution along $z$. The exact reason why is discussed in Subsection~\ref{subsec:expVStheory}. Fortunately, Eq.~\eqref{eq:M} can still be used, provided that $\sz$ is replaced by the effective $\sz$:
\begin{equation}\label{eq:szeff}
    \sigma_z^{\mathrm{eff}} = \frac{1}{2\sqrt{\pi}\int\rho^2(z)dz},
\end{equation}
\noindent where
\begin{equation}
    \rho(z)\equiv\frac{1}{\Ne}\frac{d\Ne}{dz}.
\end{equation}

Equation \eqref{eq:M} does not reveal the exact distribution for $\NPhc$, it only gives the variance $\var{\NPhc}$. However, the form of the distribution can be suggested by a simple qualitative argument when the number of longitudinal modes $M_L$ is much larger than one (for bending-magnet radiation $M_L\sim\kz\sz$, for undulator radiation $M_L\sim\kz\sz/N_u$). Indeed, in this case the total quasi-classical number of detected photons $\NPhc$ is a sum of a large number of independent random numbers of detected photons coming from small longitudinal slices of the bunch. Therefore, according to the central limit theorem, $\NPhc$ must obey a normal distribution with good accuracy. More details on the exact distribution for $\NPhc$ can be found in \cite{saldin1998statistical,kim2017synchrotron,rice1944mathematical,kim1997temporal,catravas1999measurement,huang2007review} which suggest that, in the general case of incoherent spontaneous radiation, the quasi-classical radiated power obeys Gamma statistics.

\section{Comparison with experimental data}

\subsection{Brookhaven experiment}

\begin{figure}[h!]
   \centering
   \includegraphics*[width=\columnwidth]{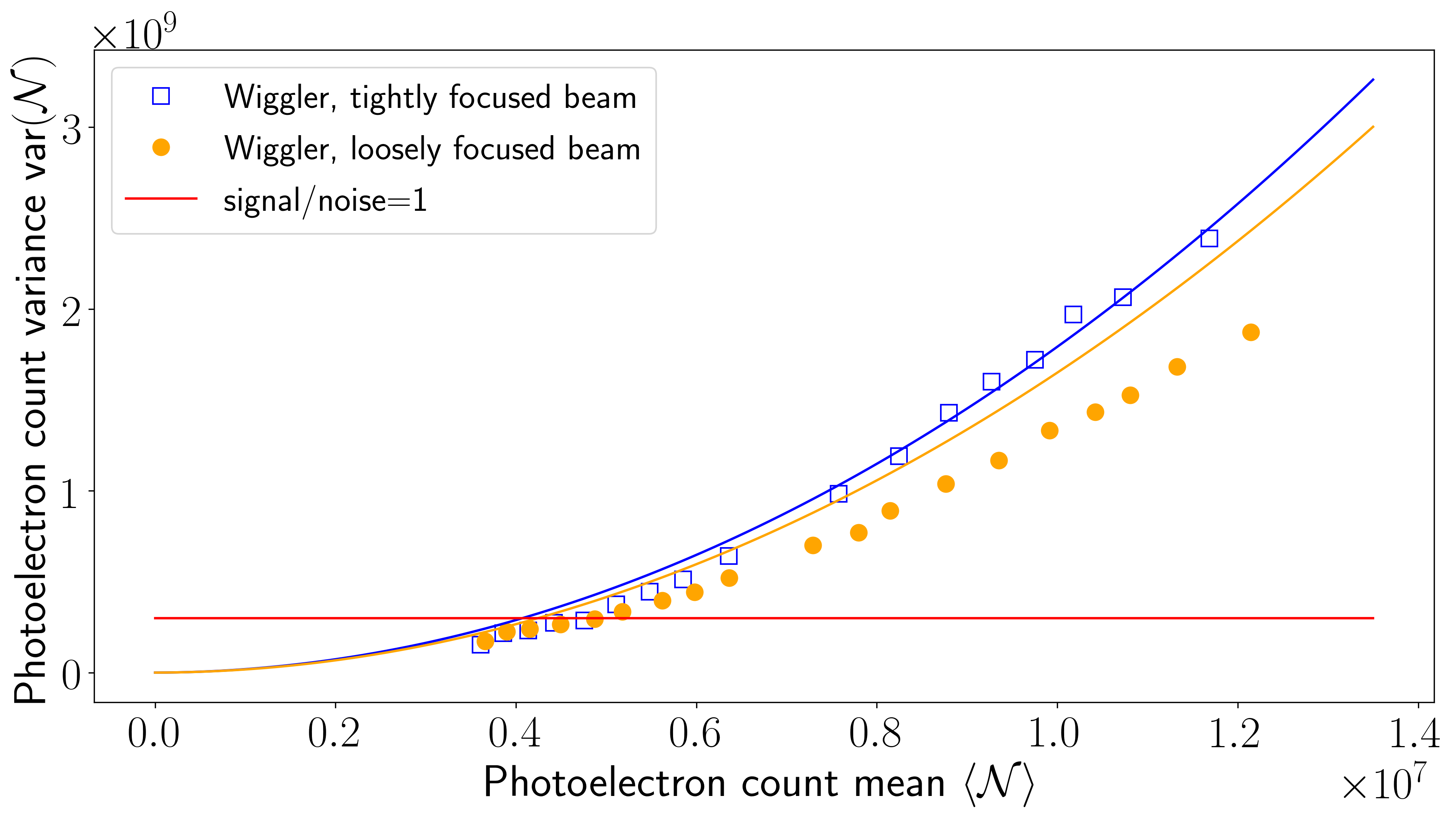}
   \caption{Experimental data from Ref.  \cite{teich1990statistical} for wiggler radiation and predictions made by our theoretical model. The noise variance ($\approx \SI{3e8}{}$) has been subtracted from the data. }
   \label{fig:bnl_data}
\end{figure}

In the early experiment at Brookhaven National Lab \cite{teich1990statistical}, the fluctuations in the wiggler and bending-magnet radiation were studied at the Brookhaven Vacuum-Ultraviolet Electron Storage Ring. The data in Fig.~\ref{fig:bnl_data} were extracted from the original paper \cite{teich1990statistical} by digitizing the plot. The scale was also changed from log-log to a linear scale. This procedure could have introduced some deviations from the original data, but the deviations are believed to be negligible. We did not attempt to compare the empirical data for bending-magnet radiation from \cite{teich1990statistical} with our theoretical model's predictions, since the authors of \cite{teich1990statistical} indicated that the data likely represented the statistical properties of the secondary photons produced in the Pyrex vacuum chamber window, rather than the statistical properties of the original bending-magnet radiation. 

The data for the wiggler radiation was collected for the fundamental harmonic, $\lambda_1=\SI{532}{nm}$. An optical interference filter with FWHM $=\SI{3.2}{nm}$ and a maximum transmission at $\lambda_1$ was used. The rms strength parameter of the wiggler was $K_w=4$, the number of periods $N_w=22.5$, the period length $\lambda_w=\SI{10}{cm}$. The electron beam energy was $\SI{650}{MeV}$. A silicon PIN photodiode was used to convert the wiggler radiation photons into photoelectrons. Two configurations of the beam optics in the vicinity of the wiggler were studied, i.e., two transverse beam profiles: a tightly focused beam and a loosely focused beam (see Ref.~\cite{teich1990statistical} and Fig.~\ref{fig:bnl_data}). The mean photoelectron count was mainly varied by using a variable neutral density filter. 

Since the values of the photoelectron count variance $\var{\N}$ for the wiggler radiation in Fig.~\ref{fig:bnl_data} are much larger than the values of photoelectron count mean $\av{\N}$, it can be argued that the quantum Poisson contribution (the first term in Eq.~\eqref{eq:varN_from_book}) is negligible in this experiment for the wiggler radiation. Therefore, according to our theoretical model the only remaining source of fluctuation is  the incoherence contribution (the second term in Eq.~\eqref{eq:varN_from_book}). We calculated it by performing numerical integration in Eq.~\eqref{eq:M} using the parameters of the electron bunch, the wiggler, and the filter, given in \cite{teich1990statistical}. The Gaussian model of the filter was used. Our theoretical model, i.e., Eq.~\eqref{eq:M}, predicted the following values for the number of coherent modes: for tightly focused beam, $M_{\mathrm{TFB}} = \SI{5.6e4}{}$; for loosely focused beam, $M_{\mathrm{LFB}} =  \SI{6.1e4}{}$. One can see in Fig.~\ref{fig:bnl_data} that our prediction for the tightly focused beam agrees with experimental points very well. 

However, the points for the loosely focused beam deviate from our prediction. In terms of the number of coherent modes the error is about $\SI{20}{\percent}$ for the loosely focused beam. It is practically impossible to find the exact reason for this disagreement now, because the measurements were taken about three decades ago, and it is difficult to reconstruct the exact conditions of the experiment. In part, this is what motivated us to carry out an independent study in IOTA.

\subsection{IOTA experiment}
The Integrable Optics Test Accelerator (IOTA), located at Fermilab's Accelerator Science and Technology (FAST) facility, is a small storage ring designed for experiments with both electron and proton beams. We refer the reader to Ref.~\cite{antipov2017iota} and Table~\ref{tab:exp_params} for the description of the ring and its parameters. In this experiment, the IOTA ring operated with electrons only.  

In IOTA, the values of the parameters of the electron bunch and of the undulator are such that the quantum and the incoherence contributions to radiation fluctuations are comparable, whereas the latter is usually dominant. Also, due to significant intrabeam scattering, the bunch dimensions strongly depend on beam current. Through Eq.~\eqref{eq:M}, the fluctuations themselves therefore acquire a complex dependence on beam current.

\subsubsection{Experimental apparatus}
\begin{figure}[!h]
\includegraphics[width=\columnwidth]{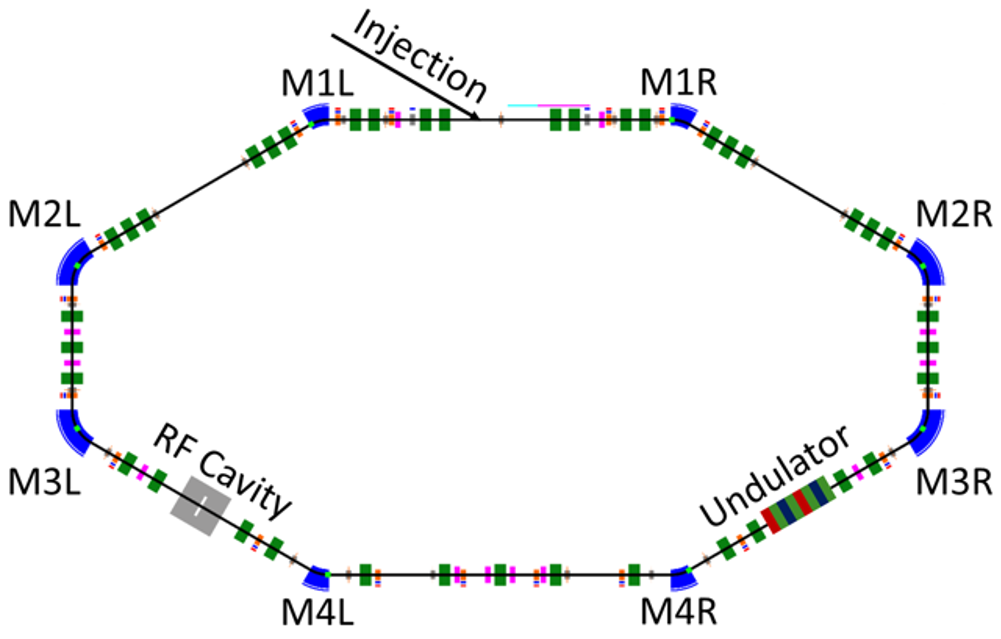}
\caption{\label{fig:iota_layout}Layout of IOTA. The circumference of the ring is 40~m.}
\end{figure}

The general layout of IOTA during the experiment is presented in Fig.~\ref{fig:iota_layout}. Beam injection takes place between M1L and M1R. The beam circulates clockwise. An undulator was inserted in the straight section between M3R and M4R. A photodetector was installed in a dark box on top of the M4R dipole magnet.

\begin{figure}[!h]
\includegraphics[width=\columnwidth]{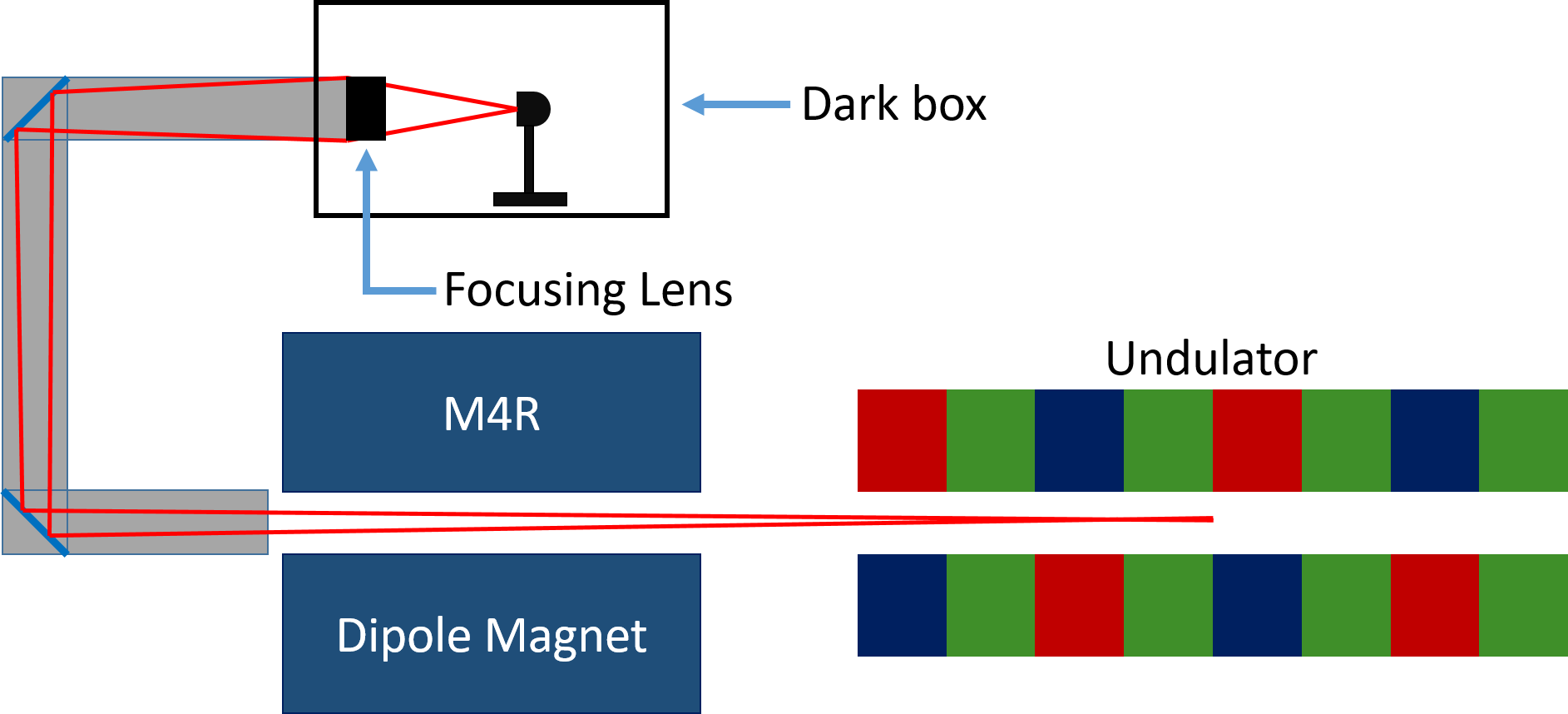}
\caption{\label{fig:periscope}Schematic of light propagation from the undulator to the photodetector (not to scale).}
\end{figure}

The light produced in the undulator was directed to the photodetector by a system of two mirrors. Then, it was focused on the sensitive area of the detector with a lens, see Fig.~\ref{fig:periscope}.

\begin{figure}[!h]
\includegraphics[width=\columnwidth]{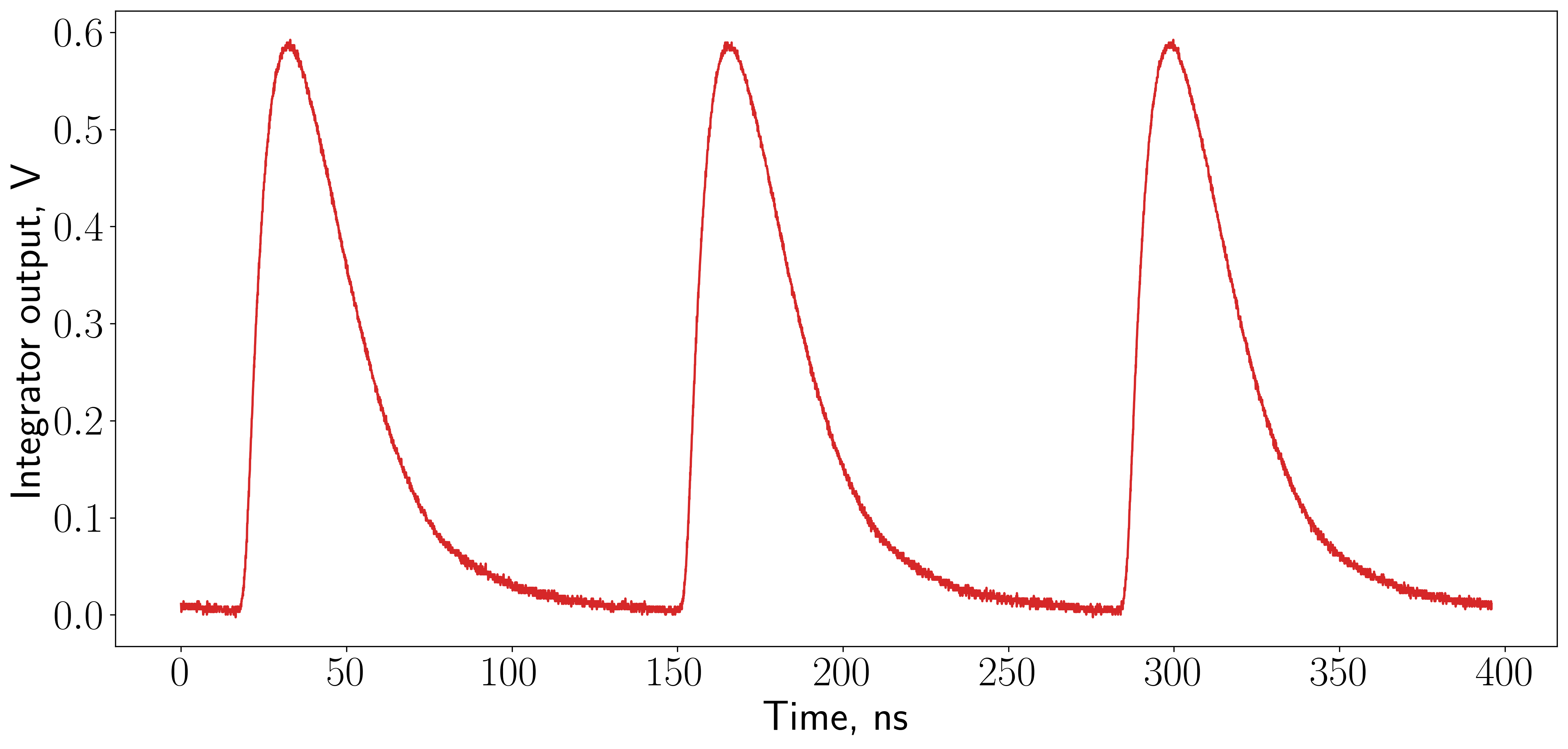}
\caption{\label{fig:integrator_output}Typical output of the photodiode current's integrator. Each pulse corresponds to one IOTA revolution.}
\end{figure}

Parameters of the undulator \cite{gottschalk2012design} installed in IOTA, along with other essential parameters of the experiment are listed in Table~\ref{tab:exp_params}.  
\begin{table}[b]
\caption{\label{tab:exp_params}%
Experimental parameters. The parameters sensitive to beam current or location are given at $\Ib=\SI{1.3}{mA}$ or in the center of the undulator.
}
\begin{ruledtabular}
\begin{tabular}{lc}
           IOTA circumference         & \SI{40}{m}       \\ 
           Revolution period   & \SI{133}{ns} \\
           Beam energy       & \SI{100}{\MeV}                \\ 
           Max average current        & \SI{4.0}{mA}              \\ 
           Emittances, $\ex$, $\ey$               & \SI{0.32}{\mu m}, \SI{31}{nm}\\
           Relative momentum spread, $\sE$                       & \SI{3.1e-4}{}\\
           Lattice functions, $\beta_x$, $\beta_y$         & \SI{1.82}{m}, \SI{1.75}{m}\\
           Dispersion, $D_x$, $D_y$                  & \SI{0.87}{m}, \SI{0}{m}\\      
           Transverse beam size, $\sx$, $\sy$  &  \SI{815}{\micro\metre}, \SI{75}{\micro\metre}\\
           Longitudinal bunch size, $\sz$   & \SI{38}{cm}\\
           Rad. damping rates,  $1/\tau_x$, $1/\tau_y$ & \SI{0.336}{s^{-1}}, \SI{0.852}{s^{-1}}\\
           $1/\tau_p$ & \SI{2.22}{s^{-1}}\\
           Undulator parameter $K_{\mathrm{u}}$       & \SI{1.0}{}                 \\
           Undulator period       & \SI{55}{mm}                 \\
           Number of undulator periods, $\Nu$       & \SI{10}{}                  \\
           Fundamental harmonic wavelength, $\lambda_1$       & \SI{1077}{\nano\metre}               \\
           Photodiode diameter       & \SI{1}{mm}                 \\
           Quantum efficiency @$\SI{1077}{\nano\metre}$      & \SI{80}{\percent}             \\
           Beam lifetime    &     $>\SI{10}{min}$
\end{tabular}
\end{ruledtabular}
\end{table}
\begin{figure}[!h]
\includegraphics[width=\columnwidth]{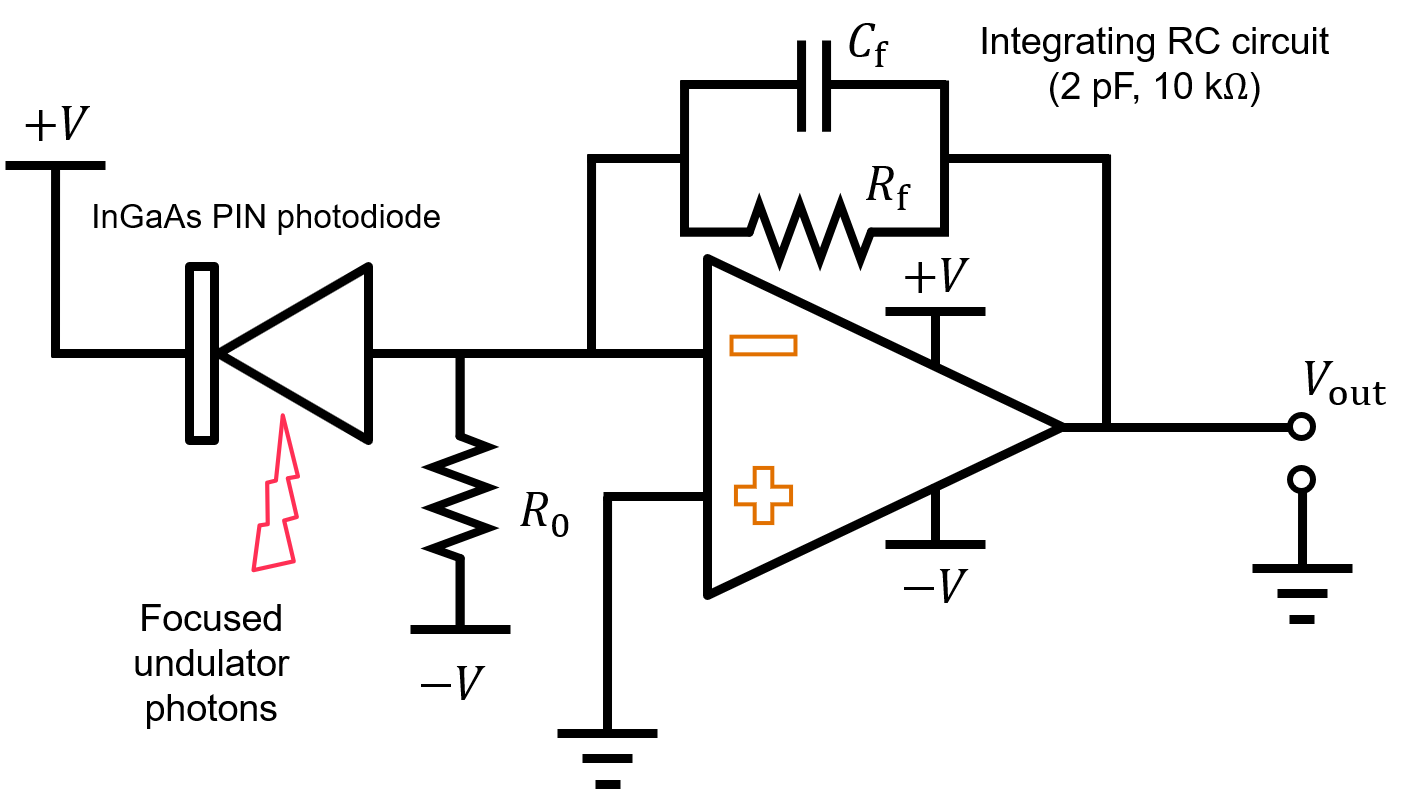}
\caption{\label{fig:integrating_circuit} A schematic of the photodetector circuit with an op-amp-based (Texas Instruments THS4304 \cite{opamp}) photocurrent integrator, $V=\SI{3.3}{V}$.}
\end{figure}
The experiment employed an InGaAs PIN photodiode (Hamamatsu G11193-10R \cite{hamamatsu_photodiode}) to convert short ($\sz/c\approx\SI{1.2}{ns}$) pulses of the undulator radiation into electric current pulses of roughly the same duration.  For this experiment, there was a single bunch in the IOTA ring, circulating with a revolution period of \SI{133}{ns}. The photodetector circuit is shown in Fig.~\ref{fig:integrating_circuit}. The photo-current pulse quickly charges a capacitor $\Cf=\SI{2}{pF}$ and then this capacitor slowly ($\Rf\Cf=\SI{20}{ns}$) discharges through a resistor, $\Rf=\SI{10}{k\ohm}$, see Fig.~\ref{fig:integrator_output}.  We also used the resistor $R_0=\SI{580}{\kilo\ohm}$ in our circuit (Fig.~\ref{fig:integrating_circuit}) to remove the offset in the output signal (about $\SI{300}{mV}$), produced by the op-amp input bias current and the photodiode leakage current.  It is important to select the value of $R_0$ as high as possible to reduce the resistor Johnson-Nyquist current noise contribution.  It should be noted that the detector circuit could be further optimized to increase the signal-to-noise ratio.  

\begin{figure}[!h]
\includegraphics[width=\columnwidth]{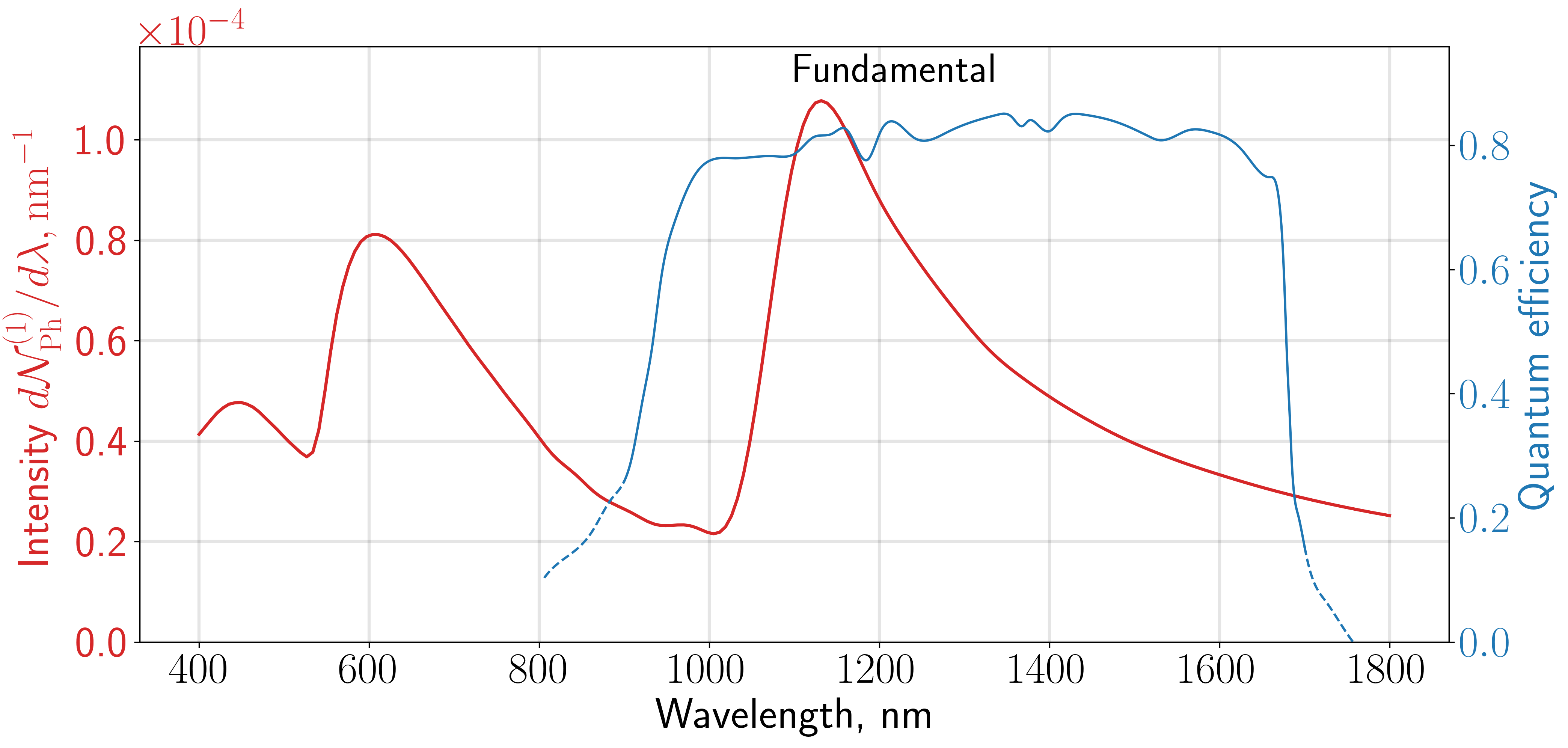}
\caption{\label{fig:undulator_spectrum}Spectral density of the number of photons emitted by a single electron in the undulator into a round aperture with $\SI{2}{in}$ diameter, located $\SI{3.5}{m}$ away from the center of the undulator. The simulation was performed in SRW \cite{chubar2013wavefront}. The quantum efficiency curve was obtained by using the photosensitivity data for the photodiode available on the Hamamatsu website \cite{hamamatsu_photodiode}.}
\end{figure}

The number of detected photons (i.e., the number of photoelectrons) is related to the voltage amplitude of the integrator output signal $A$ by
\begin{equation}\label{eq:conversion}
    \N=\frac{\Cf}{e}A,
\end{equation}
\noindent where $e$ is the electron charge. The amplitude $A$ reached values up to $\SI{1.2}{V}$ during the experiment. We studied the fundamental harmonic of the undulator radiation, $\lambda_1=\SI{1077}{nm}$. The spectrum of the fundamental was rather wide (see Fig.~\ref{fig:undulator_spectrum}) due to the small number of periods ($\Nu=\SI{10}{}$) in our undulator. The FEL gain length was $L_\mathrm{g}=\SI{4}{m}$, while the length of the undulator was only $L_u=\SI{0.6}{m}$. Therefore, we observed spontaneous undulator radiation.
We did not use a narrow spectral filter as in \cite{teich1990statistical}. To focus the radiation on the sensitive area of the photodiode ($\diameter \SI{1.0}{mm}$), we used an achromatic doublet AC508-150-C from Thorlabs ($\diameter \SI{2}{in}$) with focal length of $\SI{150}{mm}$, so that the chromatic aberration would be minimized, see Fig.~\ref{fig:periscope}. The distance between the center of the undulator and the achromatic doublet was $\SI{3.5}{m}$. An estimate for the collected angle is therefore $\theta_\mathrm{ap}=\SI{1}{in}/\SI{3.5}{m}=\SI{7}{mrad}$, which is comparable to  $1/\gamma=\SI{5}{mrad}$. We believe that the actual aperture was smaller than the estimated $\theta_\mathrm{ap}$ due to the periscope with two mirrors used to look at the undulator radiation, see Fig.~\ref{fig:periscope}. First, the mirrors ($\diameter \SI{2}{in}$) reduce vertical aperture by a factor of $\sqrt{2}$, since they are at $\SI{45}{\degree}$ to the direction of propagation of radiation field. Second, there could be some misalignment in the periscope. In our simulations in Subsec.~\ref{subsec:expVStheory} we used an elliptical aperture with horizontal semi-axis equal to $\SI{3.0}{mrad}$, and vertical semi-axis equal to $\SI{3.0}{mrad}/\sqrt{2}=\SI{2.2}{mrad}$.  However, the photon flux obtained in the experiment was about three times smaller, than what was predicted in the simulation (Fig.~\ref{fig:undulator_spectrum}). The exact reasons for the disagreement are still not well understood.

\begin{figure}[!h]
\includegraphics[width=\columnwidth]{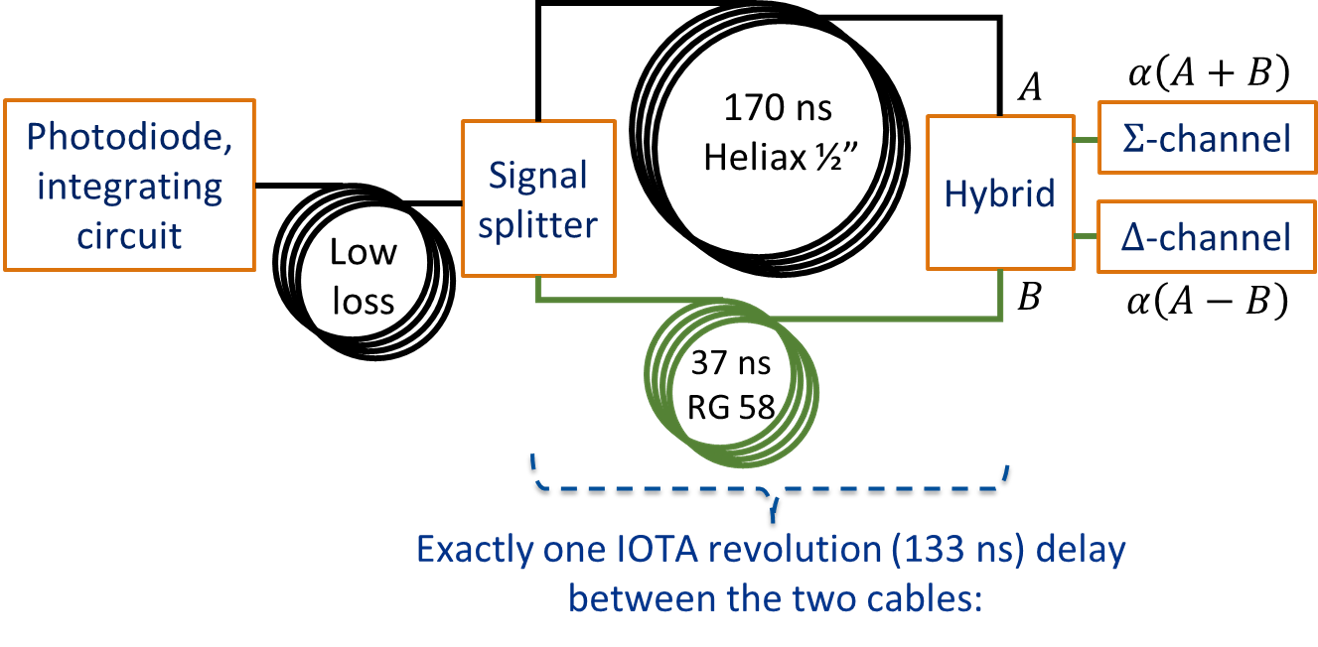}
\caption{\label{fig:comb_filter} A schematic of the comb filter, which allows us to look directly at the amplitude difference between two consecutive pulses in IOTA (\D-channel). A two-way resistive power splitter, employed in our experiment, had a 6-dB insertion loss (per output channel). The attenuation coefficient in the hybrid $\alpha$ was about 3 dB, or $\SI{0.7}{}$ in amplitudes.}
\end{figure}

It would be very hard to study small fluctuations of the amplitude ($\SI{e-4}{}-\SI{e-3}{}$ rms) with our 8 bit oscilloscope (model Rohde\&Schwarz RTO1044  4GHz 20 GSa/s) by looking directly at the integrator's output signal, see Fig.~\ref{fig:integrator_output}. To improve the sensitivity of our measurements, we use a so-called comb filter \cite{smith2010physical}, see Fig.~\ref{fig:comb_filter}. The time delay between the two arms following the signal splitter equals exactly one IOTA revolution period. An adjustable phase shifter is used to fine-tune it. The error can be made as small as a few tenths of a nanosecond. Also, the models of the cables are chosen in such a way that the losses and dispersion in the two arms are approximately the same. Then, the signals in the two arms serve as inputs to a hybrid (model MACOM H-9), whose outputs are the sum and the difference of the input signals (\S- and \D-channels, respectively). Thus, in the \D-channel we look directly at the difference between two consecutive IOTA pulses, i.e., the pulse-to-pulse fluctuation. When looking at the \D-channel with a scope, all 8 bits are used effectively. Since all the elements in the comb filter are passive, practically no noise is introduced. The cross-talk between \S- and \D-channels was $\ct\approx\SI{.7}{\percent}$, i.e., if the pulses in IOTA were perfectly identical, there would still be pulses in \D-channel with amplitude of $\ct\approx\SI{.7}{\percent}$ of the amplitude of the pulses in \S-channel. However, this effect was easy to take into account.

To take one measurement of the photoelectron count variance, we recorded 1.5-ms-long waveforms (corresponding to about 11,000 IOTA revolutions) of \D- and \S-channels with the scope at $\SI{20}{GSa/s}$. Since the beam lifetime in IOTA is longer than $\SI{10}{min}$, the change in the number of electrons in the bunch within one waveform is negligible.

\subsubsection{Setup tests and noise subtraction}
The ability of the setup to correctly measure the amplitude fluctuations in signals similar to Fig.~\ref{fig:integrator_output} was verified independently with a test light source, consisting of a laser diode with an amplifier, modulated by a pulse generator.

One difficulty that we had to overcome in the experiment in IOTA was that the noise in the \D-channel was larger than (yet of the same order as) the turn-by-turn fluctuations in the pulses. Therefore, a special noise filtering algorithm had to be developed and applied to the collected waveforms. 

The exact procedure for obtaining $\var{\N}$ and its uncertainty from the signals in the \D- and \S-channels is described in Appendix~\ref{app:noiseSubtraction}, including the case of a signal-to-noise ratio smaller than one. 

\subsubsection{Measurement results for undulator radiation in IOTA and comparison with theoretical predictions}\label{subsec:expVStheory}
\begin{figure*}[t]
\includegraphics[width=\textwidth]{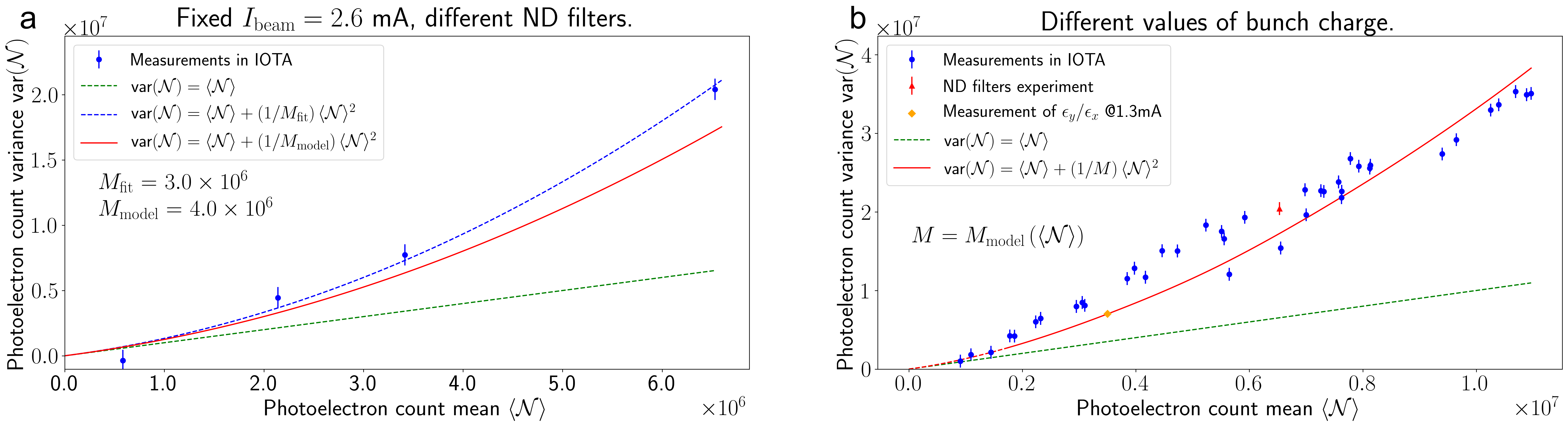}
\caption{\label{fig:iota_data}Photoelectron count variance $\var{\N}$ as a function of photoelectron count mean $\av{\N}$ for undulator radiation in IOTA. (a) $\av{\N}$ was varied by using different neutral density (ND) filters, (b) $\av{\N}$ was varied by changing the number of electrons in the bunch. The error bar is constant and equals $\SI{8.2e5}{}$, see Appendix~\ref{app:noiseSubtraction}.
}
\end{figure*}

Two sets of undulator radiation data were collected. First, measurements were taken at one fixed value of beam current, $\SI{2.6}{mA}$, and the mean photoelectron count $\av{\N}$ was changed by placing various neutral density filters in front of the detector. We employed a four-position filter slider, which was controlled remotely. The beam was re-injected for each data point, and the measurement started when the beam current decayed to $\SI{2.6}{mA}$. The plot of $\var{\N}$ as a function of $\av{\N}$ for this set of data is presented in Fig.~\ref{fig:iota_data}a. The point with maximum $\av{\N}$ represents the configuration without any filter. The blue dashed curve is a fit of the form of Eq.~\eqref{eq:varN_from_book} for the experimental points, $M_{\mathrm{fit}}=\SI{3.0e6}{}$.
In the second set of data, we did not use any neutral density filters, the mean photoelectron count $\av{\N}$ was varied by changing the electron bunch charge, see Fig.~\ref{fig:iota_data}b. The beam was re-injected several times, and every time multiple data points were collected as the current slowly decayed. The red triangle data point is the no-filter point from Fig.~\ref{fig:iota_data}a. The green dashed straight lines in Fig.~\ref{fig:iota_data}a,b represent the predicted Poisson contribution.


To make a theoretical prediction for $M$ and, consequently, for $\var{\N}$, we had to know the dimensions of the electron bunch in IOTA as a function of bunch charge. 
To be able to estimate the bunch dimensions for the range of beam current values in our experiments, we developed a theoretical model of bunch evolution, including several effects and consistent with the available experimental data.

In addition to synchrotron radiation damping and radiation diffusion, there are three main effects, determining the bunch parameters in IOTA, namely, intrabeam scattering \cite{chao2013handbook}, multiple Coulomb scattering in the background gas \cite{reiser1994theory}, and longitudinal bunch self-focusing \cite{haissinski1973exact} due to space-charge. We use the method described in \cite{nagaitsev2005intrabeam,chao2013handbook} to compute emittance growth rates associated with intrabeam scattering. Let us define the intrabeam scattering growth rates in momentum $p$, in the horizontal $x$, and in the vertical $y$-directions as
\begin{align}
\begin{aligned}
    \frac{1}{T_p}=\frac{1}{\sE^2}\frac{d\sE^2}{dt}, \quad \frac{1}{T_x}=\frac{1}{\ex}\frac{d\ex}{dt},\quad
    \frac{1}{T_y}=\frac{1}{\ey}\frac{d\ey}{dt},
\end{aligned}
\end{align}
In our simulations, we keep the longitudinal bunch size constant, $\sz=\SI{38}{cm}$. This value was determined experimentally with a wall-current monitor and it remained approximately constant for any $\Ib>\SI{0.65}{mA}$. Most likely, this is due to the above mentioned self-focusing of the electron bunch in IOTA. The self-focusing also causes some deviations from Gaussian longitudinal bunch profile. However, Eq.~\eqref{eq:M} can still be used if one substitutes $\sz$ with $\sz^{\mathrm{eff}}$, see Eq.~\eqref{eq:szeff}. Since in our model we assume one specific constant value of $\sz=\SI{38}{cm}$, it will be valid only for $\Ib>\SI{0.65}{mA}$ ($\av{\N}>\SI{1.75e6}{}$).

The horizontal and vertical betatron tunes were decoupled and linear coupling was minimized. For typical bunch dimensions, the intrabeam scattering growth rate in the $y$-direction is much smaller than the synchrotron radiation damping rate \cite{lee2018accelerator}. Therefore, vertical emittance is primarily determined by multiple Coulomb scattering in the residual gas. Moreover, this implies that vertical emittance $\ey$, and, consequently, vertical bunch size $\sy$ do not depend on beam current.

In the $x$-direction, the horizontal rms emittance at zero beam current, determined by quantum fluctuations, is $\epsilon_{x0}=\SI{3.6e-2}{\mu m}$. The contribution from  multiple Coulomb scattering is negligible compared to that from quantum fluctuations.  For $\Ib>\SI{0.65}{mA}$, both the contribution from quantum fluctuations, and the contribution from multiple Coulomb scattering are negligible compared to the intrabeam scattering growth rate (it will be shown quantitatively below), and, hence, horizontal emittance $\ex$ is defined solely by the balance between the synchrotron radiation damping rate $1/\tau_x$ (see Table~\ref{tab:exp_params}) and the intrabeam scattering growth rate ($1/T_x$):
\begin{equation}\label{eq:bal_x}
    \frac{1}{\tau_x}=\frac{1}{T_x}.
\end{equation}

In the longitudinal direction, the contribution from quantum fluctuations could in principle be neglected, since it was about one order of magnitude smaller than that from intrabeam scattering, but it was not difficult to account for it \cite{bane2002intrabeam}, therefore it was taken into consideration. Thus, the rms momentum spread $\sE$ was determined by the balance between synchrotron radiation damping on one side and intrabeam scattering growth rate and quantum fluctuations on the other side:

\begin{equation}\label{eq:bal_p}
    \frac{\sE^2}{\tau_p}=\frac{\sE^2}{T_p}+\frac{\sigma_{p0}^2}{\tau_p},
\end{equation}
\noindent where $\sigma_{p0}=\SI{8.4e-5}{}$ is the momentum spread due to synchrotron radiation alone, in the absence of intrabeam scattering; $1/\tau_p$ is the synchrotron radiation damping rate, see Table~\ref{tab:exp_params}. 

At this point in our analysis, we have three unknowns, namely, $\ex$, $\ey$, $\sE$ and only two equations, i.e., Eqs.~\eqref{eq:bal_x} and \eqref{eq:bal_p}. We do not have an equation for the $y$-direction, since we do not know the exact composition of the background gas. 

\begin{figure}[!h]
\includegraphics[width=\columnwidth]{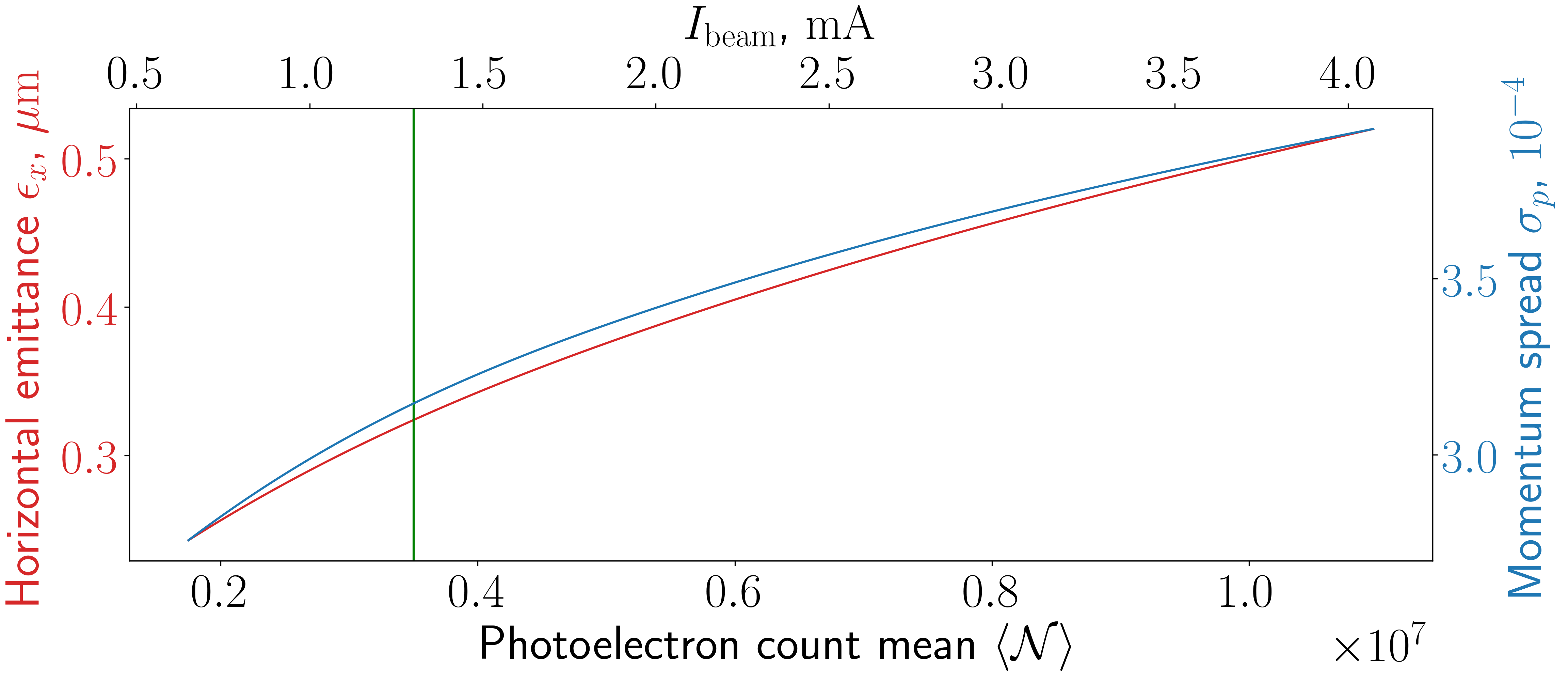}
\caption{\label{fig:ex_sgmp} Results of simulations of intrabeam scattering in IOTA. Horizontal emittance, and momentum spread in IOTA as functions of beam current $I_{\mathrm{beam}}$ and mean photoelectron count $\av{\N}$. The green vertical line indicates the beam current value $\SI{1.3}{mA}$, at which we measured the ratio of transverse emittances by looking at the images of bending-magnet radiation.}
\end{figure}

To resolve this uncertainty, we recorded several optical images of dipole-magnet synchrotron radiation from a circulating bunch at $\Ib = \SI{1.3}{mA}$. From these images and from the known betatron and dispersion functions, it was possible to determine the ratio of transverse emittances, $\ey/\ex=\SI{9.5e-2}{}$. Given this constraint, we had two equations and two unknowns at $\Ib = \SI{1.3}{mA}$. Therefore, we were able to find all the parameters of the bunch at this value of beam current, $\ex = \SI{0.32}{\mu m}$, $\ey=\SI{31}{nm}$,  $\sE=\SI{3.1e-4}{}$, $\sx=\SI{815}{\mu m}$, $\sy = \SI{75}{\mu m}$, $\sz = \SI{38}{cm}$. These are the values given in Table~\ref{tab:exp_params}. This value of the beam current is marked by an orange diamond in Fig.~\ref{fig:iota_data}b and by a green vertical line in Fig.~\ref{fig:ex_sgmp}.

As was mentioned above, we believe that the vertical emittance does not depend on the beam current. Therefore, we could use the value of $\ey=\SI{31}{nm}$ found at $\Ib=\SI{1.3}{mA}$ for other values of the beam current. Hence, at this point, at other values of the beam current we had two unknowns, $\ex$ and $\sE$, and two equations, Eqs.~\eqref{eq:bal_x} and \eqref{eq:bal_p}. Thus, we were able to compute $\ex$ and $\sE$ (see Fig.~\ref{fig:ex_sgmp}), and, consequently, all parameters of the bunch for all current values $\Ib>\SI{0.65}{mA}$. Finally,  Eq.~\eqref{eq:M} was used to compute $M$ for all values of beam current $\Ib>\SI{0.65}{mA}$, and Eq.~\eqref{eq:varN_from_book} to plot the theoretical red solid curves in Fig.~\ref{fig:iota_data}a,b.
In Fig.~\ref{fig:iota_data}b, the part of the red curve for small values of the beam current $\Ib<\SI{0.65}{mA}$ is depicted by a dashed curve, indicating that our assumption of constant $\sz$ is altered in this region. 


The experimental parameters, such as beam sizes in IOTA and the geometry of the radiation detection system, have an uncertainty of about 10\%. This translates into uncertainties in the value of the calculated parameter~$M$ of about 15\%. For a more stringent comparison with the model predictions, a better control and measurement of the beam sizes is necessary, as planned during future runs of the IOTA experimental program.

\subsubsection{Discussion}

In Fig.~\ref{fig:iota_data}a the parameter $M$ is constant. Since the measurements are taken at one specific value of beam current, the dimensions of the electron bunch are also the same during measurements with different neutral density filters. Therefore, the $M$ parameter must also stay the same. The red solid curve and the blue dashed curve in Fig.~\ref{fig:iota_data}a are parabolas of the form of Eq.~\eqref{eq:varN_from_book} with a fixed $M$.


However, in Fig.~\ref{fig:iota_data}b, the $M$ parameter changes significantly, i.e., from $M=\SI{3.5e6}{}$ at $\av{\N}=\SI{3.5e6}{}$, to $M=\SI{4.4e6}{}$ at $\av{\N} = \SI{1.1e7}{}$, due to the changes in bunch dimensions. Thus, the red solid curve in Fig.~\ref{fig:iota_data}b is no longer a parabola.

The variation of~$M$ is related to one possible application, namely, estimating 
the bunch dimensions by looking at the fluctuations in synchrotron radiation in a storage ring. Clearly, in general, the amplitude of fluctuations depends on all three bunch sizes, $\sx$, $\sy$, $\sz$. Nonetheless, in some cases additional constraints or relations may be available. For example, the ratio of transverse emittances may be known; or some bunch dimensions may be easily measurable, e.g., $\sz$ in IOTA is large and it can be readily measured with a wall-current monitor. The method of estimating the bunch dimensions by measuring the fluctuations in synchrotron radiation may be especially useful when one of the bunch dimensions is very small, and it is difficult to measure it with conventional methods. Successful measurements of longitudinal bunch size with this technique were reported in \cite{sannibale2009absolute,catravas1999measurement}. Moreover, if fluctuations data are available in a wide spectral range, the longitudinal bunch profile may be reconstructed \cite{sajaev2004measurement,sajaev2000determination}. However, it should be understood that for this method to work the radiation must be incoherent, i.e., in order to measure $\sz$, the wavelength of the radiation should be significantly smaller than $\sz$. In the opposite limit, $M_L$ is equal to one, and the fluctuations are insensitive to $\sz$.


In Fig.~\ref{fig:iota_data}a,b, one can see that the Poisson contribution (green dashed line) is comparable with the incoherence contribution (second term in Eq.~\eqref{eq:varN_from_book}). Usually the incoherence contribution is dominant \cite{kim2017synchrotron}. 
There are several reasons why the two terms in Eq.~\eqref{eq:varN_from_book} were comparable in IOTA: the small number of undulator periods $N_u=10$; a relatively small undulator parameter $K_u=1$; a relatively low beam current $\Ib<\SI{4}{mA}$ (which means small $\av{\N}$); and a relatively large $\sz=\SI{38}{cm}$ (which means large $M_L$).
To our knowledge, the experiment in IOTA is the only one where the Poisson contribution was significant in undulator radiation, as opposed to \cite{teich1990statistical,sajaev2004measurement,sajaev2000determination,catravas1999measurement}, for example. However, in bending-magnet radiation, a situation similar to ours (with a considerable Poisson term) was observed in \cite{sannibale2009absolute}.


\section{Conclusions and outlook}
We derived relations (Eqs.~\eqref{eq:varN_from_book} and \eqref{eq:M}) to predict the fluctuations $\var{\N}$ in the incoherent synchrotron radiation for a Gaussian electron bunch in undulators, wigglers and bending-magnets. The formulas properly take into account the discrete nature of light and the quantum efficiency of the detector, which is in general a function of the radiation wavelength.
A spectral filter with any transmission function can be incorporated by including a transmission function into $\qe$ in Eq.~\eqref{eq:M}. The detector acceptance can be taken into account by setting $\qe$ to zero outside of a given angular range.

The predicted variance vs. radiation intensity was compared with the empirical data from a previous experiment at Brookhaven \cite{teich1990statistical} for the case of wiggler radiation with dominant incoherence contribution 
and with new experimental data from IOTA for the case of comparable quantum 
and incoherence contributions.
In \cite{teich1990statistical}, the photoelectron count mean $\av{\N}$ was varied mainly by using different neutral density filters. In our experiment, in addition to varying $\av{\N}$ with different neutral density filters, we also varied $\av{\N}$ by changing the beam current in a wide range. The latter set of data was also used to refine the model of intrabeam scattering in IOTA.

The fact that the quantum Poisson contribution to fluctuations was significant in IOTA implied that the value of fluctuations was very small, about two orders of magnitude smaller than in \cite{teich1990statistical}. Accordingly, we introduced several critical improvements to the experimental setup that dramatically increased the measurement sensitivity. In particular, we used a comb filter with a delay equal to one IOTA revolution and a special noise filtering algorithm. 

In our present experimental configuration, the electron bunch size, undulator radiation direction, and the photodetector circuit parameters were the main sources of uncertainty. 
Future experiments in IOTA with better diagnostics and control of the beam parameters may yield more stringent comparisons between model and data and deeper insights into these phenomena.

The cleanest way to compare theory and experiment is to collect the fluctuations data at fixed beam current, as in Fig.~\ref{fig:iota_data}a. In this case, the parameter $M$ is constant, and provides an indirect measurement of the bunch dimensions. 

In the future, we plan to use a larger electron bunch (larger $M$) and consider the case when the quantum Poisson contribution is dominant, i.e., to observe the green dashed line in Fig.~\ref{fig:iota_data}a,b experimentally. This will help calibrate the photodetector circuit, since the slope of the line is determined by the capacitance $\Cf$.

As it was pointed out in \cite{sajaev2004measurement,sajaev2000determination,catravas1999measurement,sannibale2009absolute}, the fluctuations in synchrotron radiation can be used to make measurements of the bunch length on a picosecond scale, and the proof of principle experiments were successful. In IOTA, the longitudinal bunch size is relatively large, $\sz = \SI{38}{cm}$, and can be easily measured with a wall-current monitor. On the other hand, the transverse bunch size can be quite small, down to a few tens of microns, where it may be hard to measure by conventional methods. The number of coherent modes~$M$ in undulator radiation in IOTA is very sensitive to the transverse bunch size. Therefore, it may be possible to estimate the transverse bunch dimensions from the fluctuations of the number of detected photons.

\begin{acknowledgments}
We would like to thank the entire FAST/IOTA team at Fermilab for helping us with building and installing the experimental setup and taking data, especially Wayne Johnson, Mark Obrycki, and James Santucci. We also thank Greg Saewert for constructing the photodetector circuit and for providing the test light source. David Johnson and Todd Johnson provided test equipment and assisted during our detector tests. Numerous pieces of advice given by Daniil Frolov are greatly appreciated as well. Finally, A.H. is grateful to C. Pellegrini, G. Stupakov and Y. Cai (SLAC) for many in-depth physics discussions on the subject.

This research is supported by the University of Chicago and the US Department of Energy under contracts 
DE-AC02-76SF00515 and DE-AC02-06CH11357.

This manuscript has been authored by Fermi Research Alliance, LLC under Contract No. DE-AC02-07CH11359 with the U.S. Deparmentment of Energy, Office of Science, Office of High Energy Physics.

\end{acknowledgments}

\appendix
\section{Estimation of number of coherent modes}\label{app:Mestimation}
An order of magnitude estimate of the number of coherent modes $M$ can be made as the ratio of the radiation phase space volume $\Omega$ and the coherent phase space volume $\Omega_R$ \cite{sannibale2009absolute,kim2017synchrotron}:
\begin{equation}\label{eq:Mest}
    M=\frac{\Omega}{\Omega_R},
\end{equation}
\noindent where 
\begin{equation}
    \Omega_R=1/\left(2k\right)^3,
\end{equation}
\begin{multline}
    \Omega = \sqrt{1/\left(2k\right)^2+\ex^2}\sqrt{1/\left(2k\right)^2+\ey^2}\\
    \sqrt{1/\left(2k\right)^2+\epsilon_z^2},
\end{multline}
\noindent where $\epsilon_z=\sigma_z\sigma_\omega/\omega$; $k$ is the magnitude of the wave vector, $\ex$ and $\ey$ are transverse emittances, and $\sigma_\omega/\omega$ describes the width of the radiation spectrum. It is approximately equal to the inverse of the number of undualtor periods $1/\Nu$. 

In IOTA, for the fundamental harmonic of undulator radiation, $1/(2k)=\SI{8.6e-8}{m}$, $\ex=\SI{3.2e-7}{m}$, $\ey=\SI{3.1e-8}{m}$, $\epsilon_z = \SI{0.38}{m}\times1/10=\SI{3.8e-2}{m}$. With these values, Eq.~\eqref{eq:Mest} gives $M=\SI{1.8e6}{}$. This number sets the scale for the expected values of~$M$ obtained in Subsec.~\ref{subsec:expVStheory} by using Eq.~\eqref{eq:M}.

\section{System tests and noise filtering}\label{app:noiseSubtraction}

\begin{figure*}[t]
\includegraphics[width=\textwidth]{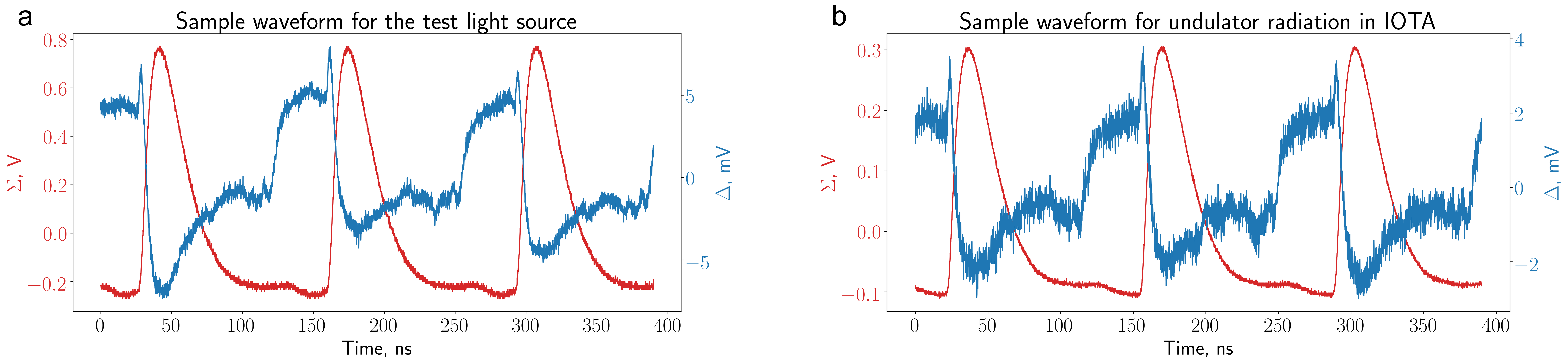}
\caption{\label{fig:sample_waveforms}(a) Sample sum (red) and difference (blue) waveforms for the test light source without any neutral density filters, $\av{\N}= \SI{1.8e7}{}$, $\var{\N}=\SI{1.1e9}{}$, signal/noise$\gg 1$; (b) Typical waveform for the undulator radiation in IOTA, $\av{\N}= \SI{0.73e7}{}$, $\var{\N}=\SI{2.3e7}{}$.  }
\end{figure*}
\begin{figure}[h]
\includegraphics[width=0.9\columnwidth]{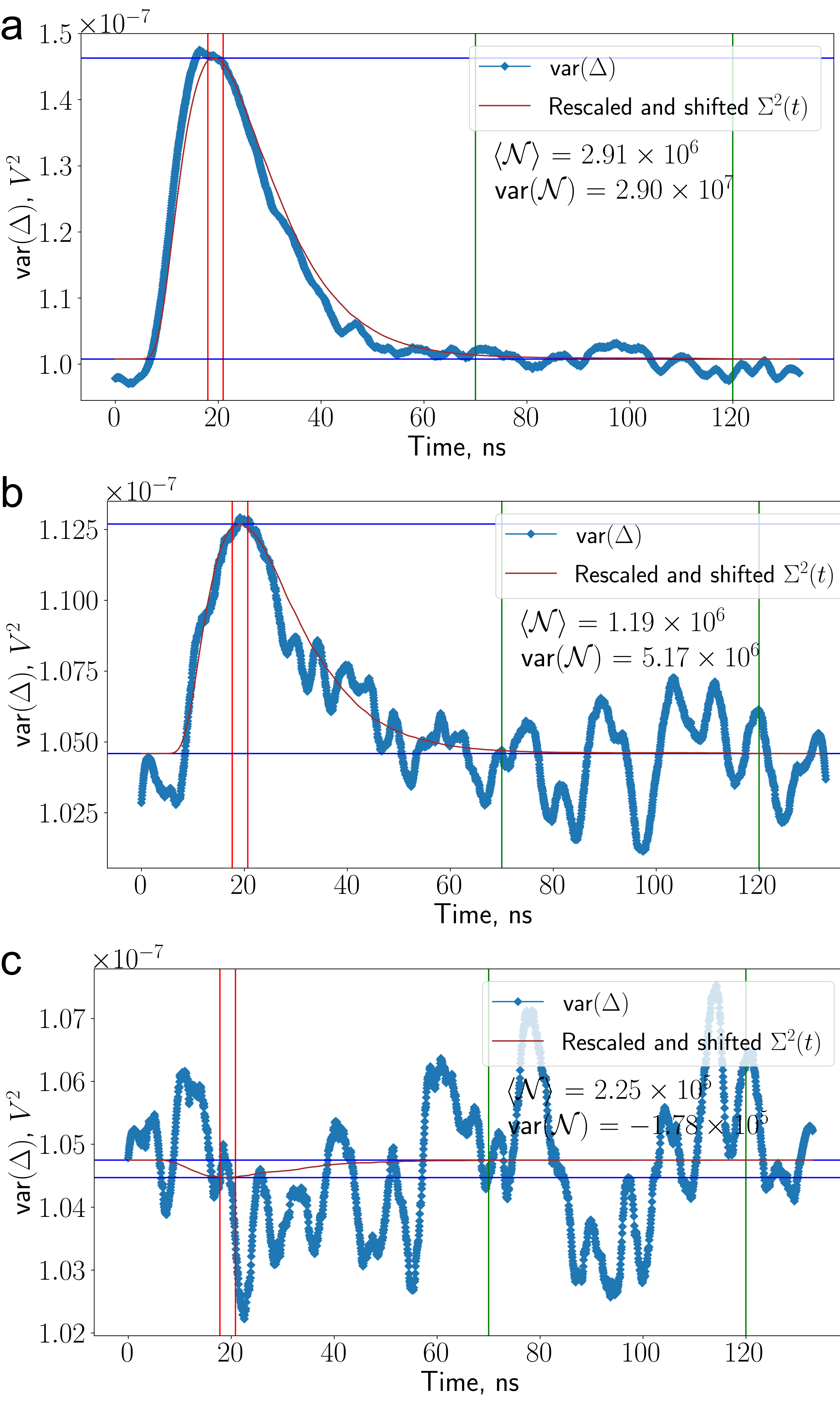}
\caption{\label{fig:noise_subtraction} Examples of the noise filtering algorithm for the test light source with different neutral density filters. (a) $\var{\Delta}$ takes the expected shape, i.e., a peak on top of a constant level; (b) Smaller fluctuations of $\Delta$, the peak is not as well defined as before; (c) Very small fluctuations of $\Delta$, the peak cannot be seen in the noise. The red curves represent the right-hand side of Eq.~\eqref{eq:Di}.}
\end{figure}

The measurement system was bench-tested with a special light source, which consisted of a fast laser diode ($\SI{1064}{nm}$) with an amplifier, modulated by a pulse generator. We also varied the mean photoelectron count (and photoelectron count variance) by placing various neutral density filters between the measurement system and the test light source. We believe that the fluctuations in the number of emitted photons in the test light source were mostly created by the jitter in the pulse generator amplitude. The unknown $\var{\N}$ of the test light source was determined experimentally in our bench tests. A typical waveform for \D- and \S-channels for the test light source without any neutral density filters is shown in Fig.~\ref{fig:sample_waveforms}a. The steps in \D- and \S-signals at $\sim\SI{80}{ns}$ after the main peaks are likely produced by signal reflections and imperfections of the hybrid, which is designed to work at frequencies between $\SI{2}{MHz}$ and $\SI{2}{GHz}$. The IOTA pulses come at $\SI{7.5}{MHz}$, which is close to the lower bandwidth limit of the hybrid. The signal at the integrator output does not show these steps, see Fig.~\ref{fig:integrator_output}.

One can see in Fig.~\ref{fig:sample_waveforms}a that the comb-filter technique works rather well. The amplitudes of the pulses in the \D-channel fluctuate significantly from pulse to pulse. And the range of these fluctuations is much larger than the noise in the \D-channel. Therefore, by analyzing these fluctuations for 11,000 periods, the relative fluctuation of photoelectron count for the test light source was quite reliably determined to be $\theta\equiv\var{\N}/\av{\N}^2=\SI{3.35e-6}{}$.

However, in the actual experiment with the undulator radiation in IOTA, the fluctuations in the pulse amplitudes in the \D-channel were smaller than for the test light source, see Fig.~\ref{fig:sample_waveforms}b. Moreover, they were smaller than the noise in the \D-channel. Therefore, it was necessary to develop a method to filter out the instrumental noise and to extract the actual fluctuations of the photoelectron count.

The idea of the method is the following.
First, we analyze the \S-channel to determine the period with which the pulses arrive with very high precision ($>7$ significant figures). In IOTA, it is the revolution time ($\SI{133}{ns}$); in the test light source, it is the period in the pulse generator, which was chosen to be $\SI{133}{ns}$ as well. Then, all the data in \S- and \D-channel are mapped onto a single period. That is, the time stamp of each sample is mapped into the remainder of a division operation between itself and the IOTA revolution period. Further, these data are binned along the time axis into about $2660$ bins, corresponding to the sampling period of the oscilloscope ($\SI{50}{ps}$). The bins are numbered by index $i = 1, \ldots, 2660$; IOTA revolutions are numbered by index $n = 1, \ldots, 11000$.
After this procedure, one has
\begin{align}
\label{eq:Sin}
\Sigma_{i,n} = V_i(1+\delta_n)+V_i(1+\delta_{n-1})+\mu \Delta_{i,n}+\xi_{i,n},\\
\label{eq:Din}
\Delta_{i,n} = V_i(1+\delta_n)-V_i(1+\delta_{n-1})+\kappa \Sigma_{i,n}+\nu_{i,n},
\end{align}
\noindent where $V_i$ represents the pulse signal from the integrator (up to a certain factor due to attenuation in signal splitter and hybrid) averaged over 11,000 IOTA revolutions; $\delta_n$ and $\delta_{n-1}$ are the relative fluctuations with respect to the average pulse signal in $n$th and $(n-1)$th turns, respectively; the parameters $\mu$ and $\kappa$ characterize the cross-talk between \S- and \D-channels; $\xi_{i,n}$ and $\nu_{i,n}$ are the noise contributions in \S- and \D-channels, respectively. Note that the noise contributions are assumed to be independent of the amplitudes of \D- and \S-signals.

For convenience, we also introduce the averaged sum signal $\Sigma_i = 2 V_i$. Hence, Eq.~\eqref{eq:Din} can be approximated as
\begin{equation}\label{eq:DinApprox}
    \Delta_{i,n} = \frac{1}{2}\Sigma_i(\delta_n-\delta_{n-1})+\kappa \Sigma_i+\nu_{i,n}.
\end{equation}

If one fixes index $i$ in Eq.~\eqref{eq:DinApprox} and takes the variance with respect to index $n$, the following relation is obtained

\begin{equation}\label{eq:Di}
    \var{\Delta_i} = \frac{1}{2}\Sigma_{i}^2\var{\delta}+\var{\nu_i}.
\end{equation}

The left-hand side of Eq.~\eqref{eq:Di} is obtained from the experimental data. Figure~\ref{fig:noise_subtraction} gives an example where the light from the test light source, significantly attenuated by neutral density filters, is studied. In this case, the signal-to-noise ratio is less than one, just like for the undulator radiation in IOTA. The horizontal axis corresponds to index $i$. However, it is represented in nanoseconds for convenience. 

On the right-hand side of Eq.~\eqref{eq:Di}, $\var{\nu_i}$ does not depend on $i$ if the noise rms amplitude is constant with time. Therefore, the contribution from noise in Fig.~\ref{fig:noise_subtraction} can be identified as the constant vertical offset. The remaining contribution is proportional to $\Sigma_{i}^2$ and looks like a peak. The value of $\var{\delta}$ can be extracted from the height of this peak using Eq.~\eqref{eq:Di}, and, therefore, $\var{\N}$ can be found.

\begin{figure*}[t]
\includegraphics[width=\textwidth]{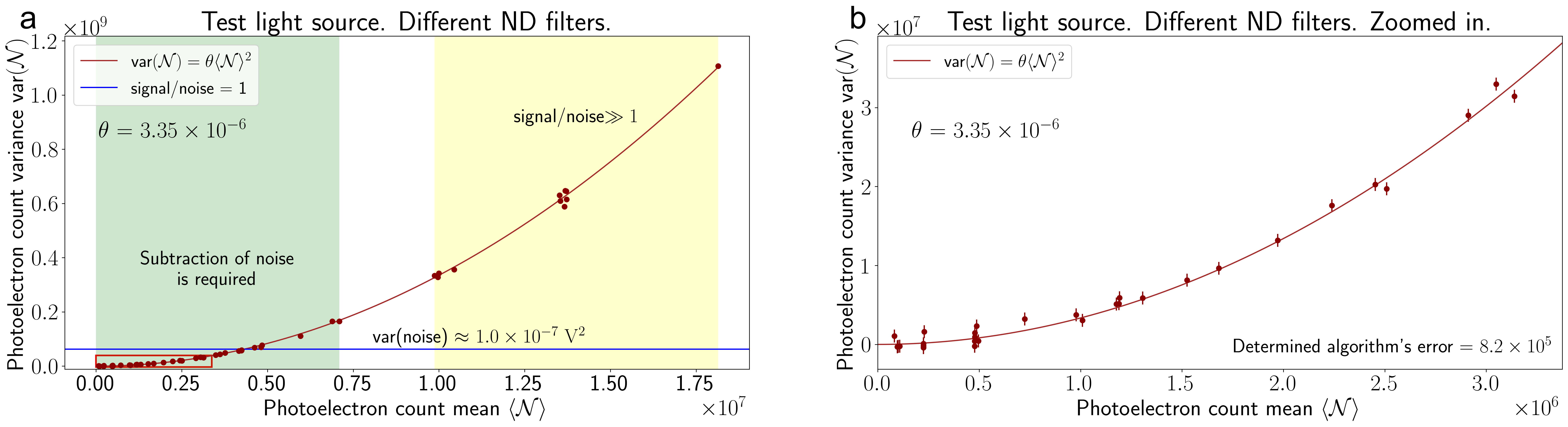}
\caption{\label{fig:test_light_source}Photoelectron count variance $\var{\N}$ as a function of photoelectron count mean $\av{\N}$ for the test light source, $\av{\N}$ was varied by using different neutral density (ND) filters.
(a) Entire considered range of $\av{\N}$; (b) The region corresponding to the actual values of $\var{\N}$ in the undulator radiation in IOTA (highlighted by the red rectangle in (a)).}
\end{figure*}

In Fig.~\ref{fig:noise_subtraction}a, we can see that the empirical $\var{\Delta_i}$ indeed takes the expected shape, i.e., a peak on top of a constant vertical offset. In Fig.~\ref{fig:noise_subtraction}b, the case of smaller fluctuations of $\Delta$ is considered, and the peak is not as well defined as in Fig.~\ref{fig:noise_subtraction}a. In Fig.~\ref{fig:noise_subtraction}c, the case of very small fluctuations of $\Delta$ is considered, and the peak cannot be distinguished from the noise. The height of the peak is calculated as the difference between the average in the region between the red vertical lines (peak region), and that in the region between the green vertical lines (noise region). Therefore, it can be seen in Fig.~\ref{fig:noise_subtraction}c, that the  method may mistakenly yield a slightly negative value for $\var{\nu_i}$ for very small fluctuations. In all three plots in Fig.~\ref{fig:noise_subtraction} the level of the constant noise variance is approximately the same, $ \SI{1.0e-7}{V^2}$, which supports the hypothesis that the noise fluctuations do not change with time or with signal level.

The observed rms noise amplitude, $\sqrt{\var{\mathrm{noise}}} = \SI{0.3}{mV}$, was analyzed using the noise model for the detector electrical schematic, Fig.~\ref{fig:integrating_circuit}, as well as the typical electrical characteristics of the photodiode~\cite{hamamatsu_photodiode} and the operational amplifier~\cite{opamp}. The three main contributions to the rms noise in the $\Delta$-channel are the following: the oscilloscope input amplifier noise, \SI{0.21}{mV}; the operational amplifier input voltage noise, \SI{0.18}{mV}; and the operational amplifier input current noise, \SI{0.037}{mV}. When added in quadrature, these three sources explain the measured noise.

We determine the height of the peak by subtracting the average of the variance in the noise region from the average in the peak region, as opposed to, for instance, fitting the curve with $\Sigma^{2}(t)$, because this procedure is more robust against small errors in the comb filter delay and jitter of the signal period.

The noise filtering method was tested with the test light source. We placed various neutral density filters in front of the test light source and measured $\var{\N}$ as a function of $\av{\N}$, see Fig.~\ref{fig:test_light_source}a.

When the light was not attenuated at all, or only slightly attenuated by neutral density filters, the signal-to-noise ratio was much larger than one and $\var{\N}$ could be determined directly by looking at the \D-channel waveform without using the noise filtering algorithm. By repeating this measurement many times for the same neutral density filter we made sure that $\var{\N}$ was stable in time for the test light source, see the group of points around $\av{N}=\SI{1.3e7}{}$ in Fig.~\ref{fig:test_light_source}a.  

When the light intensity was reduced further, the contribution from noise became substantial, and we had to use the noise filtering method to measure $\var{\N}$. However, we could also independently predict $\var{\N}$ based on the measurements of fluctuations in the region where signal/noise$\gg 1$. When quantum Poisson fluctuations can be neglected, the variance of the number of photoelectrons $\var{\N}$ scales as $\eta^2$, and the mean photoelectron count $\av{\N}$ scales as $\eta$, where $\eta$ is the attenuation factor of the neutral density filter. Therefore, the relative fluctuation remains constant
\begin{equation}\label{eq:rel_fluct}
    \theta\equiv\frac{\var{\N}}{\av{\N}^2}=\mathrm{constant}.
\end{equation}

The quantum Poisson contribution to the fluctuations in the test light source pulses could be completely neglected, because $\var{\N}$ was much larger than $\av{\N}$ in Fig.~\ref{fig:test_light_source}a,b, i.e., the fluctuations were dominated by the generator's pulse-to-pulse amplitude jitter.

From the points with signal/noise$\gg 1$ (see Fig.~\ref{fig:test_light_source}a) it was determined that $\theta=\SI{3.35e-6}{}$. The parabola defined by Eq.~\eqref{eq:rel_fluct} with this value of $\theta$ is plotted in Fig.~\ref{fig:test_light_source}a,b, solid line. It can be seen in Fig.~\ref{fig:test_light_source}a, and, especially, in zoomed-in Fig.~\ref{fig:test_light_source}b, that the points, obtained with the noise subtraction algorithm in the region where the contribution of noise is significant, closely follow the predicted parabola. In Fig.~\ref{fig:test_light_source}b, we chose a range of $\var{\N}$, similar to the one observed in the experiment with undulator radiation in IOTA.  The agreement in this plot shows that the noise subtraction algorithm works well in the regime explored by the IOTA experiment.
We estimate the uncertainty of the noise filtering algorithm from the standard deviation of the residuals between the points in Fig.~\ref{fig:test_light_source}b and the predicted parabola. This error is \SI{8.2e5}{} and it is also shown in Fig.~\ref{fig:test_light_source}b.

The noise filtering algorithm can remove noise that is independent of the amplitude of the signal and  whose rms amplitude is constant with time. Sources of such contributions to noise are, for example, oscilloscope, op-amp in the integrator, photodiode, and most external noise sources. However, it is important to emphasize that the measurements with the test light source were performed in a lab, not in the IOTA ring enclosure. Therefore, we cannot completely eliminate the possibility that in the IOTA ring enclosure the results of our measurements were affected by some sources of nonlinear time-dependent noise.


\bibliography{bibliography}

\end{document}